\documentclass[12pt,a4paper]{article}

\usepackage[utf8]{inputenc}
\usepackage{amsmath, amssymb, amsthm}
\usepackage{graphicx}
\usepackage{hyperref}
\usepackage{natbib}
\usepackage{geometry}
\usepackage{times}
\usepackage{bm}
\usepackage{algorithm}
\usepackage{algpseudocode}
\usepackage{booktabs}
\usepackage{xcolor}
\usepackage{tikz}
\usetikzlibrary{positioning, shapes, arrows, decorations.pathmorphing, calc, fit, backgrounds}
\usepackage{cleveref}
\usepackage{comment}
\usepackage{tabularx}

\usepackage{pgfplots}
\pgfplotsset{compat=1.18}
\usepackage{tikz}
\usetikzlibrary{positioning, shapes, arrows, decorations.pathreplacing, calc, fit, backgrounds}

\crefname{equation}{equation}{equations}
\crefname{figure}{figure}{figures}
\crefname{table}{table}{tables}
\crefname{section}{section}{sections}
\crefname{theorem}{theorem}{theorems}

\title{\textbf{Recursive Gaussian Processes and the Bayesian Brain}}
\author{
    Moumita Das\textsuperscript{1}, Dipanjan Ray\textsuperscript{2,3} 
    and Sourabh Bhattacharya\textsuperscript{4,\(\dagger\)} \\
    \textsuperscript{1}Indian Institute of Management, Udaipur, India \\
    \textsuperscript{2}Department of Psychology and Cognitive Sciences, Ashoka University, Haryana, India \\
    \textsuperscript{3}Koita Centre for Digital Health at Ashoka (KCDH-A), Haryana, India \\
    \textsuperscript{4}Interdisciplinary Statistical Research Unit, Indian Statistical Institute, Kolkata, India \\
    \(\dagger\)Corresponding author: \texttt{bhsourabh@gmail.com}
}

\date{}

\begin{document}

\maketitle

\begin{abstract}
Predictive coding stands as the dominant theoretical framework for understanding cortical computation, positing that the brain implements hierarchical Bayesian inference through the reciprocal exchange of top-down predictions and bottom-up prediction errors. The canonical microcircuit provides the neuroanatomical substrate for these computations, with distinct laminar populations playing specialized roles. Concurrently, advances in Bayesian deep learning have sought to embed similar principles of hierarchical representation and uncertainty quantification into artificial neural networks. In this work, we establish a formal connection between predictive coding and our recently proposed Recursive Gaussian Process (RGP) framework. The RGP architecture employs a single Gaussian process $g(\cdot,\cdot)$ that takes both layer index and input value as arguments, enabling layer-specific functions $g^{(t)}(\cdot) = g(t,\cdot)$ to share statistical strength through a carefully designed covariance structure while retaining the flexibility for functional specialization. We demonstrate that RGPs naturally implement key predictive coding principles: hierarchical Bayesian inference, uncertainty propagation, prevention of representational collapse, and recursive refinement. We map RGP components onto the canonical microcircuit—the single GP corresponding to shared genetic and developmental constraints across cortical areas, and spike-and-slab variable selection implementing precision-weighted prediction error. Drawing on the free energy principle, we show that RGP inference minimizes variational free energy, thereby establishing a formal link between Bayesian mechanics and neuronal dynamics. We contrast RGP with traditional deep learning and Bayesian neural networks, demonstrating why conventional approaches fall short of implementing predictive coding while RGPs naturally embody its core principles. This synthesis positions RGPs as both a powerful computational tool and a candidate model for the brain's hierarchical predictive machinery, offering testable predictions about laminar-specific dynamics, precision-weighting mechanisms, and spectral asymmetries between feedforward and feedback processing.
    \\[2mm]
\end{abstract}

\newpage
\tableofcontents

% =============================================================================
% Diagram 1: The Predictive Coding Hierarchy with Layer-Specific KL Divergence
% =============================================================================
%\begin{figure}[t]
\begin{figure}[htpb]
\centering
\begin{tikzpicture}[
    node distance = 2.5cm and 3.5cm,
    prediction/.style = {rectangle, draw=blue!60, fill=blue!20, thick, minimum width=2.5cm, minimum height=1.2cm, align=center, font=\small},
    error/.style = {rectangle, draw=red!60, fill=red!20, thick, minimum width=2.5cm, minimum height=1.2cm, align=center, font=\small},
    sensory/.style = {rectangle, draw=green!60, fill=green!20, thick, minimum width=2.5cm, minimum height=1.2cm, align=center, font=\small},
    kl/.style = {rectangle, draw=orange!60, fill=orange!10, thick, minimum width=2.5cm, minimum height=1.2cm, align=center, font=\small},
    arrow/.style = {->, >=stealth, thick},
    doublearrow/.style = {<->, >=stealth, thick}
]

% Level 3 (Highest)
\node[prediction] (pred3) {Predictions\\L3};
\node[error, below=1.5cm of pred3] (err3) {Prediction\\Errors L3};
\node[kl, right=4cm of err3] (kl3) {$\mathfrak{h}_3$};

% Level 2
\node[prediction, below left=2.5cm and 1cm of err3] (pred2) {Predictions\\L2};
\node[error, below=1.5cm of pred2] (err2) {Prediction\\Errors L2};
\node[kl, right=4cm of err2] (kl2) {$\mathfrak{h}_2$};

% Level 1
\node[prediction, below left=2.5cm and 1cm of err2] (pred1) {Predictions\\L1};
\node[error, below=1.5cm of pred1] (err1) {Prediction\\Errors L1};
\node[kl, right=4cm of err1] (kl1) {$\mathfrak{h}_1$};

% Sensory Input
\node[sensory, below=1.5cm of err1] (sensory) {Sensory\\Input};

% Connections between levels
\draw[arrow] (err3) -- node[right, pos=0.4, font=\small] {Feedforward} (pred2);
\draw[arrow] (pred2) -- node[left, pos=0.6, font=\small] {Feedback} (err3);
\draw[arrow] (err2) -- node[right, pos=0.4, font=\small] {Feedforward} (pred1);
\draw[arrow] (pred1) -- node[left, pos=0.6, font=\small] {Feedback} (err2);
\draw[arrow] (err1) -- (sensory);
\draw[arrow] (sensory) -- (err1);

% Intra-level connections
\draw[doublearrow] (pred3) -- (err3);
\draw[doublearrow] (pred2) -- (err2);
\draw[doublearrow] (pred1) -- (err1);

% KL connections
\draw[arrow, dashed] (err3) -- (kl3);
\draw[arrow, dashed] (err2) -- (kl2);
\draw[arrow, dashed] (err1) -- (kl1);

% Labels
\node[above, font=\small] at (pred3.north) {Higher Cortical Areas};
\node[below, font=\small] at (sensory.south) {Sensory Periphery};

% KL formula box - centered below
\node[align=left, font=\footnotesize, draw=gray!30, fill=gray!5, rounded corners, text width=12cm] at (0,-8) {
    \textbf{Layer-specific KL divergence rates} $\mathfrak{h}_t$: In predictive coding, each layer minimizes its local prediction error. The KL divergence rate at layer $t$ measures the precision-weighted prediction error between the top-down prediction $\eta_t$ and the bottom-up signal $\eta_{t-1}$:
    \[
    \mathfrak{h}_t = \log\left(\frac{\sigma_t}{\sigma_{t-1}}\right) - \frac{1}{2} + \frac{\sigma_{t-1}^2}{2\sigma_t^2} + \frac{1}{2\sigma_t^2}\mathbb{E}\big[(\eta_t - \eta_{t-1})^2\big]
    \]
    This is analogous to the global KL divergence rate $\mathfrak{h}(\boldsymbol{\theta})$ in \eqref{eq:rgp_kl_rate_explicit}, applied locally at each hierarchical level. The precision $1/(2\sigma_t^2)$ weights the prediction error based on its reliability, implementing attention-like gain control \citep{feldman2010}.
};

\end{tikzpicture}
\caption{The hierarchical architecture of predictive coding with layer-specific KL divergence rates $\mathfrak{h}_t$. Each level minimizes its local precision-weighted prediction error, contributing to global free energy minimization.}
\label{fig:predictive-coding-hierarchy}
\end{figure}
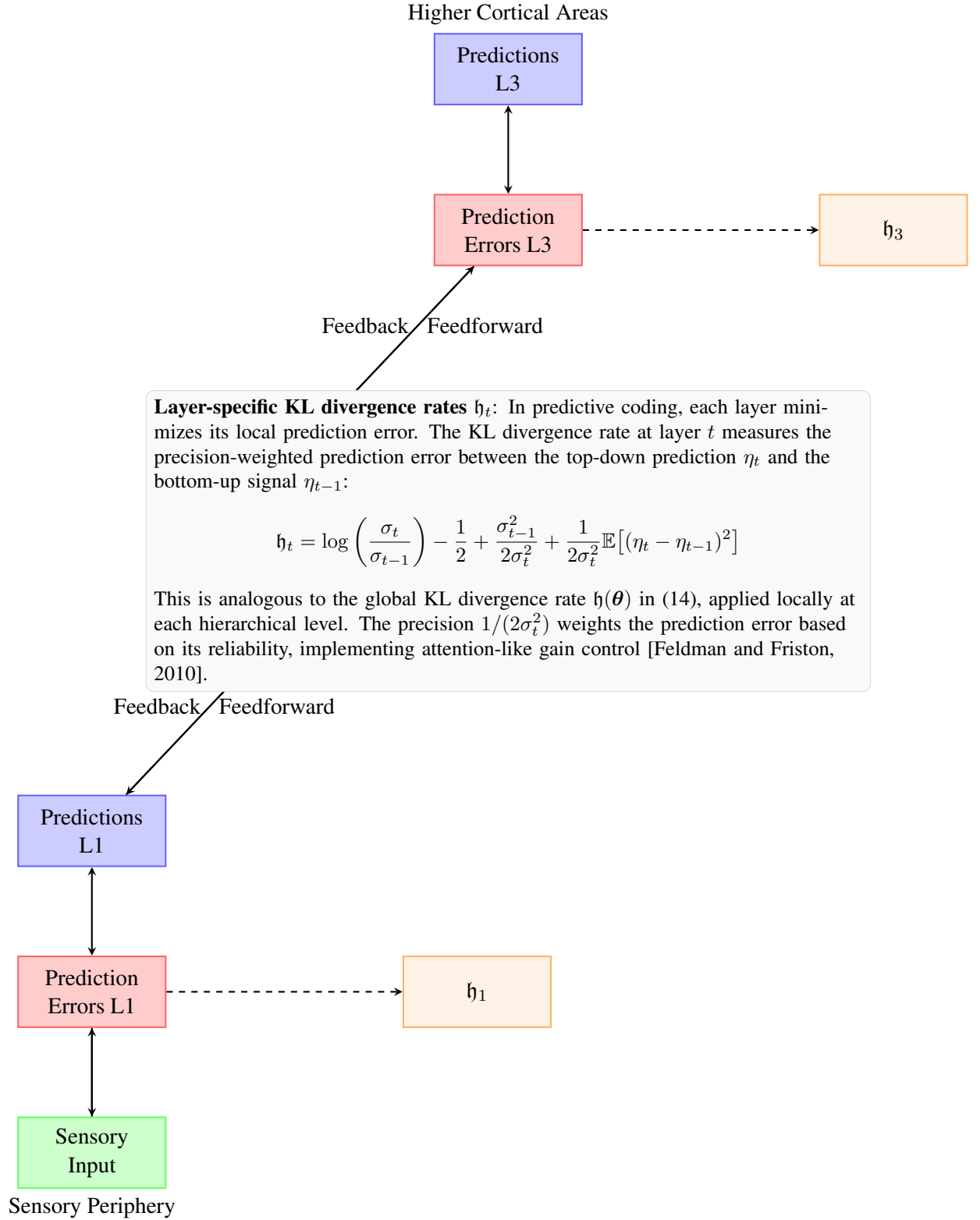

% =============================================================================
% Diagram 2: The Canonical Microcircuit
% =============================================================================
%\begin{figure}[t]
\begin{figure}[htpb]
\centering
\begin{tikzpicture}[
    node distance = 2cm and 3cm,
    superficial/.style = {rectangle, draw=purple!60, fill=purple!20, thick, minimum width=3cm, minimum height=1.2cm, align=center, font=\small},
    deep/.style = {rectangle, draw=blue!60, fill=blue!20, thick, minimum width=3cm, minimum height=1.2cm, align=center, font=\small},
    granular/.style = {rectangle, draw=orange!60, fill=orange!20, thick, minimum width=3cm, minimum height=1.2cm, align=center, font=\small},
    inhibitory/.style = {rectangle, draw=red!60, fill=red!20, thick, minimum width=3cm, minimum height=1.2cm, align=center, font=\small},
    arrow/.style = {->, >=stealth, thick}
]

% Layer labels on left
\node[align=right, font=\small] at (-4.5,0) {Layer 2/3 (Superficial)};
\node[align=right, font=\small] at (-4.5,-3) {Layer 4 (Granular)};
\node[align=right, font=\small] at (-4.5,-6) {Layer 5/6 (Deep)};

% Cell populations
\node[superficial] (sp) at (0,0) {Superficial\\Pyramidal};
\node[inhibitory] (si) at (4,0) {Inhibitory\\Interneurons};
\node[granular] (gr) at (2,-3) {Granular\\Cells};
\node[deep] (dp) at (0,-6) {Deep\\Pyramidal};
\node[inhibitory] (di) at (4,-6) {Inhibitory\\Interneurons};

% External connections
\node[font=\small] at (-2,1.8) {From Higher Areas};
\node[font=\small] at (6,1.8) {To Higher Areas};
\node[font=\small] at (-2,-7.8) {To Lower Areas};
\node[font=\small] at (6,-7.8) {From Lower Areas};

% Connections
\draw[arrow] (-2,1.5) -- (sp.north west);
\draw[arrow] (si.north east) -- (6,1.5);
\draw[arrow] (sp) -- (si);
\draw[arrow] (si) -- (sp);
\draw[arrow] (sp) -- (gr);
\draw[arrow] (gr) -- (sp);
\draw[arrow] (gr) -- (dp);
\draw[arrow] (dp) -- (gr);
\draw[arrow] (dp) -- (di);
\draw[arrow] (di) -- (dp);
\draw[arrow] (dp.south west) -- (-2,-7.5);
\draw[arrow] (6,-7.5) -- (di.south east);

% Functional labels
\node[align=center, font=\small, text width=3cm] at (6.5,0) {Precision-Weighting\\(Attention)};
\node[align=center, font=\small, text width=3cm] at (6.5,-3) {Mismatch\\Computation};
\node[align=center, font=\small, text width=3cm] at (6.5,-6) {Prediction\\Generation};

% Bounding box
\draw[dashed, gray] (-3,1.2) rectangle (5.5,-7.2);

\end{tikzpicture}
\caption{The canonical microcircuit for predictive coding (adapted from \citet{bastos2012}). Superficial pyramidal cells encode prediction errors and originate feedforward connections. Deep pyramidal cells encode predictions and originate feedback connections. Granular layer neurons mediate the comparison between predictions and sensory data. Inhibitory interneurons implement precision-weighting.}
\label{fig:canonical-microcircuit}
\end{figure}
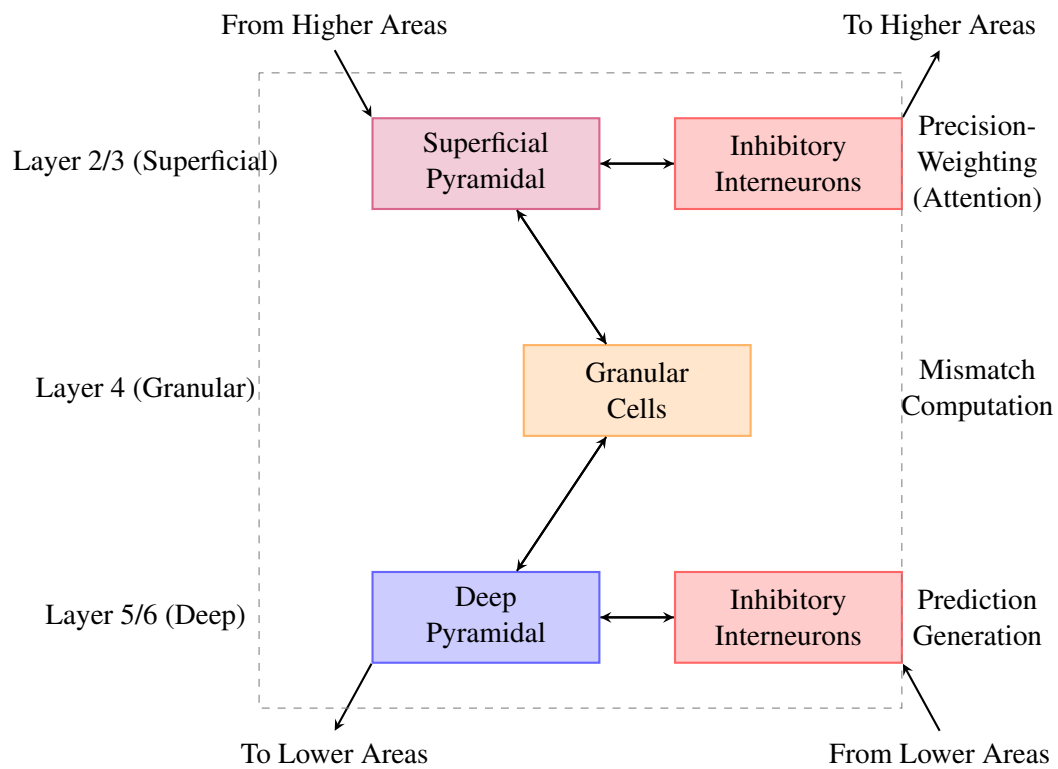

\section{Introduction}
\label{sec:introduction}

The quest to understand the computational principles underlying brain function has long been intertwined with advances in machine learning and statistical inference. In recent years, this intersection has crystallized into a dedicated research program---NeuroAI---that seeks to leverage neuroscience insights to inspire artificial intelligence innovations while using computational models to test theories of neural computation \citep{zunger2024, hassabis2017}. At the heart of this enterprise lies predictive coding, arguably the most comprehensive theoretical framework for understanding perception, action, and learning in biological systems \citep{friston2005, rao1999}.

Predictive coding posits that the brain is not a passive recipient of sensory information but rather an active inference engine that continuously generates predictions about its sensory inputs and updates these predictions based on discrepancies between expectations and actual observations \citep{clark2013}. This framework conceptualizes the cortex as a hierarchical Bayesian inference machine: higher cortical areas generate top-down predictions of neural activity in lower areas, while lower areas compute prediction errors that are propagated upward to refine higher-level representations \citep{mumford1992}. Through this iterative exchange, the brain minimizes prediction error and arrives at a coherent interpretation of the sensory world. \Cref{fig:predictive-coding-hierarchy} illustrates this hierarchical message-passing architecture.

The neuroanatomical basis for these computations has been extensively characterized in the canonical microcircuit literature. \citet{bastos2012} synthesized anatomical and physiological evidence to propose a detailed mapping between predictive coding computations and the laminar architecture of cortex. In their framework, superficial pyramidal cells (layers 2/3) encode prediction errors and originate feedforward connections, while deep pyramidal cells (layers 5/6) encode predictions and originate feedback connections. Granular layer (layer 4) neurons receive feedforward input and mediate the comparison between predictions and sensory data. Inhibitory interneurons at each layer implement precision-weighting mechanisms that modulate the gain of prediction error units based on their reliability \citep{feldman2010}. \Cref{fig:canonical-microcircuit} provides a schematic of this canonical microcircuit.

Remarkably, this neuroscientific framework finds a natural mathematical counterpart in Bayesian deep learning, particularly in models that combine hierarchical structure with principled uncertainty quantification. 
%\textcolor{cyan}
{Recent developments in machine learning have transformed predictive coding from a theory of cortical computation into a general computational framework capable of training deep architectures. \citet{millidge2022predictivecodingfuturedeep} demonstrate that predictive coding networks can approximate, and under certain conditions exactly reproduce, the gradient updates computed by backpropagation while relying only on local computations and iterative error correction. This result is particularly important from a NeuroAI perspective because it suggests that hierarchical Bayesian inference and efficient learning need not be mutually exclusive. Rather, predictive coding may provide a biologically plausible route to achieving the representational power traditionally associated with deep learning. Our recently proposed Recursive Gaussian Process (RGP) framework \citep{bhattacharya2025} represents a novel contribution to this landscape. RGP framework aligns naturally with this perspective: instead of propagating gradients through independent layers, information is recursively refined through hierarchical Bayesian inference and uncertainty propagation, resembling the iterative error-correction dynamics central to predictive coding.}
 RGPs embed a single Gaussian process prior within a hierarchical architecture inspired by deep neural networks, creating a class of Bayesian nonparametric models that preserve expressivity while providing rigorous uncertainty quantification. Crucially, the RGP architecture was explicitly designed to overcome the degeneracy problems that plague conventional deep Gaussian processes (DGPs)---where repeated composition of GP layers leads to pathological loss of variability and representational collapse \citep{duvenaud2014, dunlop2018}. Unlike standard DGPs that compose independent GPs, RGP uses a single GP $g(\cdot,\cdot)$ that takes both layer index $t$ and input value as arguments, with layer-specific functions $g^{(t)}(\cdot) = g(t,\cdot)$ sharing statistical strength through a carefully designed covariance structure that allows the model to learn the appropriate degree of dependence across layers.

% =============================================================================
% Diagram 3: Deep Gaussian Process Degeneracy (FIXED)
% =============================================================================
%\begin{figure}[t]
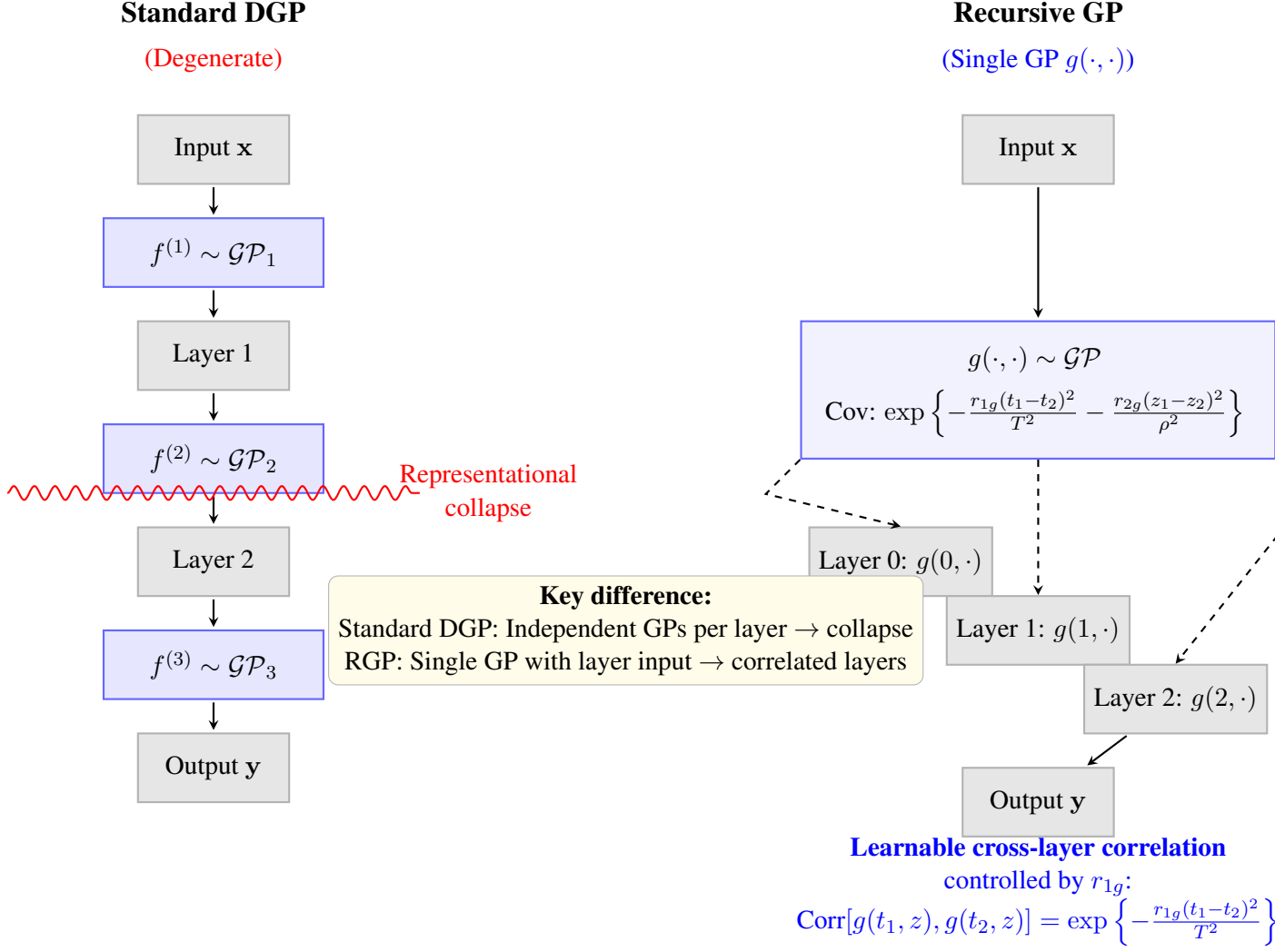
\begin{figure}[htpb]
\centering
\begin{tikzpicture}[
    node distance = 2cm and 3.5cm,
    layer/.style = {rectangle, draw=gray!60, fill=gray!20, thick, minimum width=2.2cm, minimum height=1cm, align=center, font=\small},
    function/.style = {rectangle, draw=blue!60, fill=blue!10, thick, minimum width=3.2cm, minimum height=1cm, align=center, font=\small},
    arrow/.style = {->, >=stealth, thick, shorten >=1pt, shorten <=1pt}
]

% Left side: Standard DGP (degenerate)
\begin{scope}[xshift=-6cm]
\node[align=center, font=\bf] at (0,5) {Standard DGP};
\node[align=center, font=\small, text=red] at (0,4.3) {(Degenerate)};

\node[layer] (inputL) at (0,3) {Input $\mathbf{x}$};
\node[function] (f1) at (0,1.5) {$f^{(1)} \sim \mathcal{GP}_1$};
\node[layer] (h1L) at (0,0) {Layer 1};
\node[function] (f2) at (0,-1.5) {$f^{(2)} \sim \mathcal{GP}_2$};
\node[layer] (h2L) at (0,-3) {Layer 2};
\node[function] (f3) at (0,-4.5) {$f^{(3)} \sim \mathcal{GP}_3$};
\node[layer] (outputL) at (0,-6) {Output $\mathbf{y}$};

\draw[arrow] (inputL) -- (f1);
\draw[arrow] (f1) -- (h1L);
\draw[arrow] (h1L) -- (f2);
\draw[arrow] (f2) -- (h2L);
\draw[arrow] (h2L) -- (f3);
\draw[arrow] (f3) -- (outputL);

% Collapse indication
\draw[decorate, decoration={snake, amplitude=1mm, segment length=3mm}, red, thick] (-3,-2) -- (3,-2);
\node[red, font=\small, align=center] at (4,-2) {Representational\\collapse};
\end{scope}

% Right side: RGP (non-degenerate)
\begin{scope}[xshift=6cm]
\node[align=center, font=\bf] at (0,5) {Recursive GP};
\node[align=center, font=\small, text=blue] at (0,4.3) {(Single GP $g(\cdot,\cdot)$)};

\node[layer] (inputR) at (0,3) {Input $\mathbf{x}$};

% Single GP box - larger and centered
\node[function, minimum height=2cm, minimum width=5cm, fill=blue!5] (g) at (0,-0.5) {
    \begin{tabular}{c}
        $g(\cdot,\cdot) \sim \mathcal{GP}$\\[2mm]
        Cov: $\exp\left\{-\frac{r_{1g}(t_1-t_2)^2}{T^2} - \frac{r_{2g}(z_1-z_2)^2}{\rho^2}\right\}$
    \end{tabular}
};

\node[layer] (h0R) at (-2,-3) {Layer 0: $g(0,\cdot)$};
\node[layer] (h1R) at (0,-4) {Layer 1: $g(1,\cdot)$};
\node[layer] (h2R) at (2,-5) {Layer 2: $g(2,\cdot)$};
\node[layer] (outputR) at (0,-6.5) {Output $\mathbf{y}$};

\draw[arrow] (inputR) -- (g);

% Dashed arrows from GP to each layer
\draw[dashed, ->, >=stealth, thick] (g.south west) -- ++(-0.5,-0.5) -- (h0R.north);
\draw[dashed, ->, >=stealth, thick] (g.south) -- (h1R.north);
\draw[dashed, ->, >=stealth, thick] (g.south east) -- ++(0.5,-0.5) -- (h2R.north);

\draw[arrow] (h2R) -- (outputR);

% Dependence indication - placed at bottom
\node[blue, font=\small, align=center] at (0,-7.8) {
    \textbf{Learnable cross-layer correlation}\\
    controlled by $r_{1g}$:\\
    $\text{Corr}[g(t_1,z),g(t_2,z)] = \exp\left\{-\frac{r_{1g}(t_1-t_2)^2}{T^2}\right\}$
};
\end{scope}

% Center annotation
\node[draw=black!30, fill=yellow!10, rounded corners, align=center, font=\small] at (0,-4) {
    \textbf{Key difference:}\\
    Standard DGP: Independent GPs per layer $\rightarrow$ collapse\\
    RGP: Single GP with layer input $\rightarrow$ correlated layers
};

\end{tikzpicture}
\caption{Comparison between standard Deep Gaussian Processes (left) and RGP (right). Standard DGPs suffer from representational collapse as independent GPs are repeatedly composed. RGPs prevent collapse by using a single GP $g(\cdot,\cdot)$ with layer-index input, with correlation across layers controlled by $r_{1g}$. The covariance structure allows layers to share statistical strength while maintaining the ability to specialize.}
\label{fig:dgp-degeneracy}
\end{figure}

The free energy principle \citep{friston2010} provides a unifying theoretical framework that connects predictive coding to fundamental physical principles. According to this principle, any self-organizing system that maintains its integrity over time must minimize its variational free energy---a quantity that bounds the surprise of sensory observations. Predictive coding emerges as a specific process theory for how the brain might implement free energy minimization through hierarchical message passing. \citet{friston2009} demonstrated that under the Laplace approximation and mean-field assumptions, free energy minimization reduces to precision-weighted prediction error minimization, providing a formal foundation for linking neuronal dynamics to Bayesian inference.

The epistemological status of Bayesian models in neuroscience has been critically examined by \citet{colombo2012}, who distinguish between instrumentalist and realist interpretations. They argue that psychophysical evidence showing that humans behave as optimal Bayesian observers does not, by itself, warrant the conclusion that the brain implements Bayesian inference at the mechanistic level. Rather, such evidence supports an instrumentalist interpretation: Bayesian models are useful tools for predicting and systematizing behavioral data, but they may not correspond to actual neural mechanisms. Establishing a realist interpretation requires linking algorithmic-level descriptions to implementational-level mechanisms---precisely the kind of evidence that the canonical microcircuit literature provides.

A critical question arises: can traditional deep learning models be said to implement predictive coding? While deep neural networks share some architectural features with cortical hierarchies, such as layered organization and increasingly abstract representations, they differ fundamentally in their computational principles. Standard deep networks are trained via backpropagation, a procedure that lacks clear biological plausibility and does not naturally incorporate uncertainty quantification or Bayesian inference. Moreover, the feedforward nature of most deep networks, with only limited feedback connections, stands in contrast to the recurrent message-passing architecture central to predictive coding. \Cref{fig:dgp-degeneracy} illustrates one key difference: while standard deep Gaussian processes suffer from representational collapse due to composition of independent GPs, RGPs maintain expressivity through a single GP with layer-index input and learnable cross-layer correlation.

In this paper, we argue that the RGP framework provides a compelling computational instantiation of predictive coding principles that bridges algorithmic-level descriptions and implementational-level mechanisms. We demonstrate that the hierarchical structure of RGPs, their Bayesian foundations, and their recursive inference mechanisms align remarkably well with the core tenets of predictive coding theory as articulated in the canonical microcircuit literature. Moreover, we show that the innovations introduced in RGPs---a single GP with layer-index input, spike-and-slab variable selection, and fully Bayesian MCMC inference---address fundamental limitations that have previously hindered the biological plausibility of deep Gaussian process models.

%\textcolor{cyan}
{Recent work on hierarchical active inference has emphasized that predictive processing must extend beyond perception to support planning and decision making across multiple temporal and spatial scales. \citet{rangarajan2026hierarchical} demonstrate that hierarchical active inference can be implemented through multiscale representations that support efficient planning while minimizing variational free energy. This perspective suggests that hierarchical Bayesian models should not merely encode sensory predictions but should also maintain uncertainty-aware representations across levels of abstraction. The RGP architecture naturally provides such a mechanism: each recursive layer represents a different level of abstraction while maintaining coherent uncertainty estimates through a shared Gaussian process prior. Consequently, RGP may be interpreted not only as a model of predictive perception but also as a computational substrate for hierarchical active inference.}

The paper is structured as follows. \Cref{sec:canonical-microcircuit} reviews the canonical microcircuit model of predictive coding. \Cref{sec:rgp-architecture} introduces the RGP framework and its key innovations. \Cref{sec:connecting-rgp} establishes formal connections between RGPs and predictive coding. \Cref{sec:free-energy} draws on the free energy principle to show that RGP inference minimizes variational free energy. \Cref{sec:epistemological} critically examines the epistemological status of Bayesian models. \Cref{sec:simulation-results} discusses simulation results from the original RGP paper through the lens of predictive coding. \Cref{sec:comparison-traditional} provides a comprehensive comparison between traditional deep learning and RGP. \Cref{sec:comparison-bayesian} presents a comparative analysis of Bayesian neural networks, Bayesian deep learning, and RGP. \Cref{sec:parallel-computation} examines the parallel computation architecture of RGPs and its neurobiological correspondence. \Cref{sec:discussion} discusses implications and outlines testable predictions. \Cref{sec:conclusion} concludes.

% =============================================================================
% Diagram 4: RGP Architecture (COMPLETELY REDESIGNED - NO OVERLAPS)
% =============================================================================
%\begin{figure}[t]
\begin{figure}[htpb]
\centering
\begin{tikzpicture}[
    node distance = 2cm and 4cm,
    input/.style = {rectangle, draw=green!60, fill=green!20, thick, minimum width=2.2cm, minimum height=1cm, align=center, font=\small},
    gp/.style = {rectangle, draw=blue!60, fill=blue!5, thick, minimum width=5.5cm, minimum height=2.8cm, align=center, font=\small},
    hidden/.style = {rectangle, draw=purple!60, fill=purple!20, thick, minimum width=2.5cm, minimum height=1cm, align=center, font=\small},
    weight/.style = {rectangle, draw=orange!60, fill=orange!10, thick, minimum width=2cm, minimum height=0.8cm, align=center, font=\small},
    output/.style = {rectangle, draw=red!60, fill=red!20, thick, minimum width=2.2cm, minimum height=1cm, align=center, font=\small},
    arrow/.style = {->, >=stealth, thick, shorten >=3pt, shorten <=3pt},
    dashedarrow/.style = {->, >=stealth, thick, dashed, shorten >=3pt, shorten <=3pt}
]

% Place the GP box at the top center
\node[gp] (g) at (5,2) {
    \begin{tabular}{c}
        \textbf{Single Gaussian Process} \\[2pt]
        $g(\cdot,\cdot) \sim \mathcal{GP}$ \\[4pt]
        $c_g(z_1(t_1), z_2(t_2)) = $ \\
        $\exp\left\{-\frac{r_{1g}(t_1-t_2)^2}{T^2} - \frac{r_{2g}(z_1-z_2)^2}{\rho^2}\right\}$
    \end{tabular}
};

% Input layer - left side
\node[input] (x) at (0,-1) {Input $\mathbf{x}_i$};

% Layer 0 components - positioned vertically
\node[weight] (w0) at (0,-2.5) {$\mathbf{w}^{(0)}$};
\node[hidden] (h0) at (0,-4) {$\mathbf{h}_i^{(0)}$\\$g(0,\cdot)$};

% Layer 1 components - shifted right
\node[weight] (w1) at (3,-3.5) {$\mathbf{w}^{(1)}$};
\node[hidden] (h1) at (3,-5) {$\mathbf{h}_i^{(1)}$\\$g(1,\cdot)$};

% Layer 2 components - further right
\node[weight] (w2) at (6,-4) {$\mathbf{w}^{(2)}$};
\node[hidden] (h2) at (6,-5.5) {$\mathbf{h}_i^{(2)}$\\$g(2,\cdot)$};

% Layer T components - rightmost
\node[weight] (wT) at (9,-3) {$\mathbf{w}^{(T)}$};
\node[hidden] (hT) at (9,-4.5) {$\mathbf{h}_i^{(T)}$\\$g(T,\cdot)$};

% Output section - bottom right
\node[weight] (wtilde) at (9,-6.5) {$\tilde{\mathbf{w}}$};
\node[output] (y) at (9,-8) {Output $\mathbf{y}_{ij}$};
\node at (7.5,-7.5) {$\varepsilon_{ij}$};

% Draw feedforward connections
\draw[arrow] (x) -- (w0);
\draw[arrow] (w0) -- (h0);
\draw[arrow] (h0.east) -- ++(1,0) |- (w1.west);
\draw[arrow] (w1) -- (h1);
\draw[arrow] (h1.east) -- ++(1.5,0) |- (w2.west);
\draw[arrow] (w2) -- (h2);
\draw[arrow] (h2.east) -- ++(2,0) |- (wT.west);
\draw[arrow] (wT) -- (hT);
\draw[arrow] (hT.south) -- (wtilde);
\draw[arrow] (wtilde) -- (y);
\draw[arrow] (7.5,-7.5) -- (y);

% Draw GP connections (dashed arrows from GP to each hidden layer)
\draw[dashedarrow] (g.south west) -- ++(-0.5,-0.5) -- (h0.north east);
\draw[dashedarrow] (g.south) -- ++(0,-0.8) -| (h1.north);
\draw[dashedarrow] (g.south east) -- ++(0.5,-1) -| (h2.north);
\draw[dashedarrow] (g.east) -- ++(2,0) |- (hT.north);

% Parameter explanations - placed at bottom in two columns
\node[draw=gray!30, fill=gray!5, rounded corners, align=left, font=\footnotesize, text width=4.5cm] at (2,-10) {
    \textbf{Spike-and-Slab Priors:}\\
    $w_j^{(t)} \sim \tilde{p}\mathcal{N}(0,\sigma_1^2) + (1-\tilde{p})\mathcal{N}(0,\sigma_2^2)$\\
    \quad $\tilde{p}=0.5$, $\sigma_1^2=0.1$, $\sigma_2^2=1$
};

\node[draw=blue!30, fill=blue!5, rounded corners, align=left, font=\footnotesize, text width=4.5cm] at (8,-10) {
    \textbf{Key Parameters:}\\
    $r_{1g}$: Cross-layer correlation\\
    $r_{2g}$: Input-space correlation\\
    $\rho$: Length scale\\
    $T$: Total layers
};

% Legend
\node[align=center, font=\tiny] at (5,-11) {
    \tikz\draw[arrow] (0,0) -- (0.3,0); = Feedforward \qquad
    \tikz\draw[dashedarrow] (0,0) -- (0.3,0); = GP evaluation
};

\end{tikzpicture}
\caption{The Recursive Gaussian Process architecture. A single Gaussian process $g(\cdot,\cdot)$ takes both layer index $t$ and input value as arguments, generating layer-specific functions $g^{(t)}(\cdot) = g(t,\cdot)$. The covariance structure allows the model to learn the degree of dependence across layers through $r_{1g}$. Spike-and-slab priors on weights $\mathbf{w}^{(t)}$ implement automatic variable selection and precision-weighting.}
\label{fig:rgp-architecture}
\end{figure}
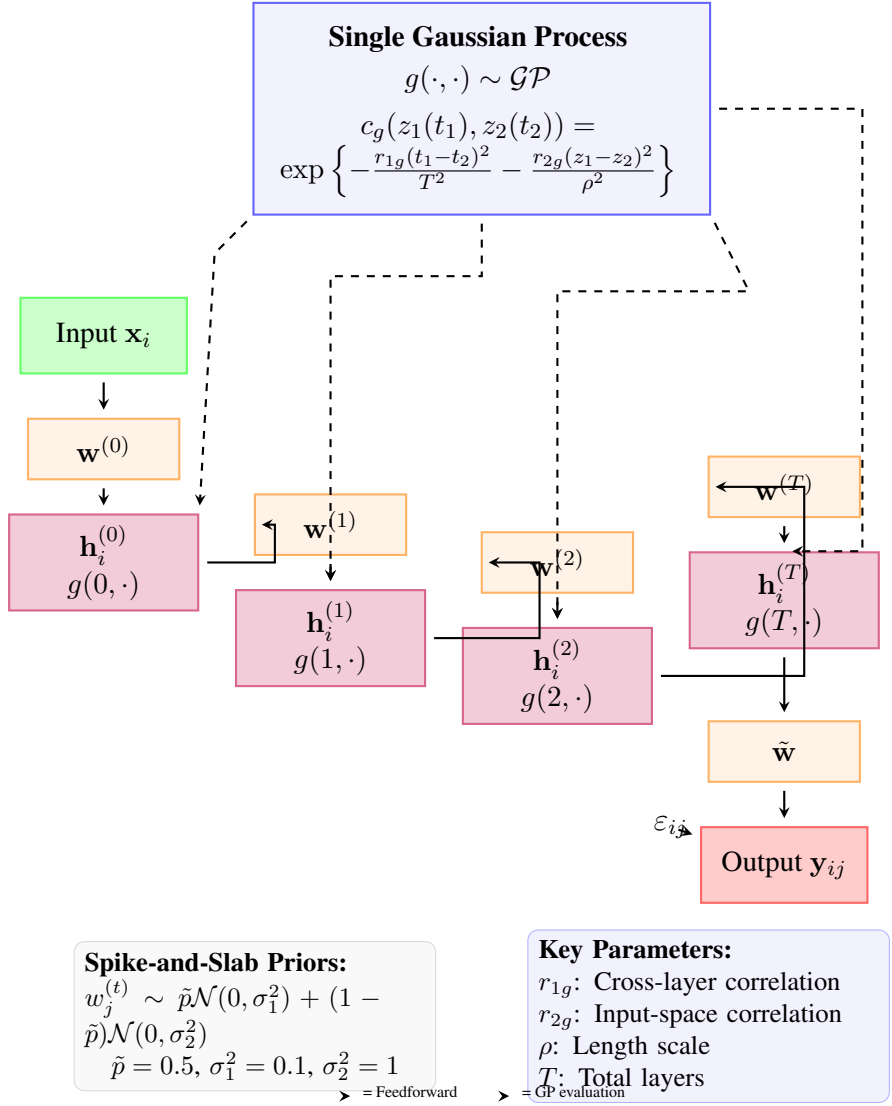

\section{The Canonical Microcircuit for Predictive Coding}
\label{sec:canonical-microcircuit}

The notion of a canonical microcircuit---a stereotypical pattern of connectivity repeated across cortical areas---has deep roots in neuroanatomy. %\textcolor{cyan}
{A recurring question in systems neuroscience is whether a common computational architecture underlies the diverse functions of the cerebral cortex.} \citet{douglas1991} proposed that the cat visual cortex contains a canonical circuit comprising superficial and deep pyramidal cells with a common pool of inhibitory neurons, all receiving thalamic drive and interconnected in a characteristic pattern. This circuit could amplify transient inputs, maintain sustained activity, and balance excitation and inhibition---computations essential for cortical function. Subsequent work has refined and extended this model, incorporating detailed quantitative data on connection probabilities and strengths \citep{thomson2002, thomson2003}.

\citet{bastos2012} synthesized anatomical, physiological, and computational evidence to propose a canonical microcircuit specifically tailored to predictive coding. Their model assigns distinct computational roles to different laminar populations based on their connectivity patterns and physiological properties. Superficial pyramidal cells in layers 2/3 are proposed to encode prediction errors, originating feedforward connections that target the granular layer of higher cortical areas. Deep pyramidal cells in layers 5/6 encode predictions, originating feedback connections that target the superficial and deep layers of lower cortical areas. The laminar specificity of these connections ensures that predictions flow downward while prediction errors flow upward, creating a hierarchical message-passing architecture as illustrated in Figure \ref{fig:predictive-coding-hierarchy}. %\textcolor{cyan}
{This laminar segregation naturally separates top-down generative signals from bottom-up sensory evidence, reducing interference between the two streams while enabling iterative Bayesian updating across cortical hierarchies.}

Granular layer neurons in layer 4 receive feedforward input and are positioned to mediate the comparison between predictions and sensory data. Inhibitory interneurons have been proposed to implement precision-weighting mechanisms that modulate the gain of prediction error units based on their reliability. \citet{feldman2010} proposed that attention corresponds to the optimization of precision parameters, with increased precision amplifying the influence of reliable prediction errors and decreased precision attenuating unreliable signals.

Spectral asymmetries between superficial and deep layers provide additional evidence for the predictive coding interpretation. Superficial layers exhibit stronger gamma-band (30-80 Hz) oscillations, while deep layers exhibit stronger alpha/beta-band (8-30 Hz) oscillations \citep{roopun2006, maier2010, buffalo2011}. \citet{bastos2012} interpret this asymmetry in terms of the differential dynamics of prediction errors and predictions: prediction errors, being transient and stimulus-driven, align with faster gamma rhythms, while predictions, being sustained and context-dependent, align with slower alpha/beta rhythms.

The functional logic of this architecture becomes apparent when considered in light of hierarchical generative models. \citet{mumford1992}  %\textcolor{cyan}
{anticipated many of the core principles underlying predictive coding decades before formal Bayesian formulations became widespread.} proposing that cortico-cortical loops implement a form of analysis-by-synthesis in which higher areas generate hypotheses about the causes of sensory input and lower areas test these hypotheses against actual data.

Empirical support for this framework comes from multiple sources. Repetition suppression---the reduction in neural responses to repeated stimuli---can be explained as the progressive refinement of predictions, with predictable stimuli generating less prediction error \citep{summerfield2008, todorovic2011}. Mismatch negativity---the enhanced response to deviant stimuli---reflects the prediction error generated when sensory input violates expectations \citep{garrido2009, wacongne2011}. Omission-related responses---neural activity evoked by the absence of an expected stimulus---provide particularly striking evidence, demonstrating that the brain generates predictions even in the absence of sensory input \citep{nordby1994, yabe1997}. %\textcolor{cyan}
{Collectively, these phenomena suggest that cortical activity reflects not merely sensory encoding but continuous comparison between expected and observed inputs.}

%\textcolor{cyan}
{From a computational perspective, predictive coding can be viewed as an iterative message-passing algorithm on a hierarchical probabilistic graphical model. Each cortical level maintains latent beliefs about hidden causes of sensory inputs, while bidirectional interactions iteratively minimize variational free energy by exchanging predictions and prediction errors. This interpretation establishes a close relationship between predictive coding, Bayesian inference, and modern probabilistic machine learning.}

\section{Recursive Gaussian Processes: Architecture and Innovations}
\label{sec:rgp-architecture}

%\textcolor{cyan}
{Although predictive coding provides a compelling normative framework for hierarchical inference, specifying flexible nonlinear generative models remains challenging. Gaussian Processes offer an attractive solution because they define distributions over functions, naturally quantify uncertainty, and avoid committing to fixed parametric forms. The natural way to extend these properties to a hierarchical, recursive architecture is to stack GPs into layers---yet, as we show below, this natural extension fails in a way that motivates the RGP framework. \Cref{fig:rgp-architecture} provides a visual representation of the complete RGP architecture.
}

\subsection{Motivation: Overcoming Deep Gaussian Process Degeneracy}
\label{subsec:overcome_degeneracy}

DGPs were introduced as a natural Bayesian counterpart to deep neural networks, combining hierarchical representations with principled uncertainty quantification \citep{damianou2013}. A DGP composes multiple independent GP layers, with the output of one layer serving as the input to the next:
\begin{equation*}
f^{(1)}(x) \sim \mathcal{GP}(0, k^{(1)}(x,x')), \quad f^{(2)}(z) \sim \mathcal{GP}(0, k^{(2)}(z,z')), \quad \ldots
\end{equation*}

However, repeated composition of independent GP layers leads to degeneracy and loss of expressivity. \citet{neal1996} observed that infinitely deep compositions converge to trivial limits---either constant functions or white noise processes. \citet{damianou2013} noted severe practical difficulties in maintaining expressivity as the number of layers increases. \citet{duvenaud2014} demonstrated that when squared exponential kernels are stacked, the distribution of outputs rapidly contracts, erasing input distinctions. \citet{dunlop2018} provided a rigorous mathematical analysis, showing that composition of smooth kernels leads to pathological loss of variability.

%\textcolor{cyan}
{This failure mode is not just a mathematical curiosity; it is biologically implausible. If cortical hierarchies suffered from analogous collapse, higher areas would become functionally superfluous---yet the brain maintains rich, informative representations across multiple hierarchical levels \citep{dicarlo2012}. This suggests that evolution has discovered architectural principles that prevent representational collapse, and it is this principle that RGP is designed to formalize.
}

%If cortical hierarchies suffered from analogous collapse, higher areas would become functionally superfluous---yet the brain maintains rich, informative representations across multiple hierarchical levels \citep{dicarlo2012}. This suggests that evolution has discovered architectural principles that prevent representational collapse.

\subsection{The RGP Solution: A Single GP with Layer-Index Input}

Rather than composing independent GPs, RGP uses a {\it single Gaussian process} that takes both the layer index and input value as arguments %\textcolor{cyan}
{ so that layer-specific behavior emerges from one shared function rather than a chain of separate ones.} For a supervised learning setting with input $\mathbf{x} = (x_1,\ldots,x_p)^{\prime}$ and $T$ hidden layers:

\begin{align}
h_{ij}^{(0)} &= g(0, \mathbf{w}^{(0)\prime}\mathbf{x}_i + b_j^{(0)}), \quad j = 1,\ldots,k_0 \label{eq:rgp_layer0}\\
h_{ij}^{(t)} &= g(t, \mathbf{w}^{(t)\prime}\mathbf{h}_i^{(t-1)} + b_j^{(t)}), \quad j = 1,\ldots,k_t, \quad t = 1,\ldots,T \label{eq:rgp_layert}\\
y_{ij} &= f(\tilde{\mathbf{w}}^{\prime}\mathbf{h}_i^{(T)} + \tilde{b}_j) + \varepsilon_{ij}, \quad j = 1,\ldots,q \label{eq:rgp_output}
\end{align}

where $g(\cdot,\cdot)$ is the single Gaussian process, and $f(\cdot)$ is a separate GP governing the output mapping \citep{bhattacharya2025}. The layer-specific functions used above are simply evaluations of this single GP at different layer indices, $g^{(t)}(\cdot) = g(t,\cdot)$, so that all layers draw on shared statistical structure while remaining free to specialize.

That sharing is controlled by the prior placed on $g$, which has a structured covariance function capturing dependencies across both layer index and input value:

\begin{align}
g(\cdot,\cdot) &\sim \mathcal{GP}(\boldsymbol{\eta}(\cdot,\cdot)^{\prime}\boldsymbol{\beta}_g, \sigma_g^2 c_g(\cdot,\cdot)) \label{eq:gp_prior}\\
	c_g((t_1,z_1), (t_2,z_2)) &= \exp\left\{-\frac{r_{1g}(t_1 - t_2)^2}{T^2} - \frac{r_{2g}(z_1 - z_2)^2}{\rho^2}\right\} \label{eq:gp_cov}
\end{align}

where $r_{1g}, r_{2g} > 0$ control the similarity of the GP across layers and input values respectively. The parameter $r_{1g}$ determines how correlated the functions are at different layers: when $r_{1g}$ is small, $g(t_1,\cdot)$ and $g(t_2,\cdot)$ can be nearly independent, allowing layers to develop distinct representations; when $r_{1g}$ is large, the functions are strongly correlated, enforcing similarity.

\subsection{The Critical Role of $r_{1g}$: Learning Layer Dependence}

The covariance between layers $t_1$ and $t_2$ at the same input value $z$ is:

\begin{equation}
\text{Cov}[g(t_1, z), g(t_2, z)] = \sigma_g^2 \exp\left\{-\frac{r_{1g}(t_1 - t_2)^2}{T^2}\right\}
\label{eq:layer_cov}
\end{equation}
%\textcolor{magenta}{
%When $r_{1g}$ is learned to be small, this covariance decays slowly with layer distance, allowing layers to develop independent representations---essential for preventing representational collapse. When $r_{1g}$ is learned to be large, layers are strongly coupled, beneficial when they share structure. This learnable dependence has a natural biological interpretation: cortical areas share genetic and developmental constraints causing similar but not identical functional properties, with the degree of similarity modifiable through learning and experience.}

%\textcolor{cyan}
{The value of $r_{1g}$---learned from data rather than fixed by the modeler---determines where RGP sits on a spectrum between independent layers and fully coupled layers. A small learned $r_{1g}$ lets this covariance decay slowly with layer distance, so layers can develop distinct representations; this is the mechanism that prevents the representational collapse afflicting conventional DGPs. A large learned $r_{1g}$ instead couples layers strongly, which is advantageous when layers genuinely share structure. Unlike DGPs, where independence (or its absence) is baked into the architecture, RGP lets the data determine the right degree of layer dependence. This learnable dependence also has a natural biological interpretation: cortical areas share genetic and developmental constraints that produce similar, but not identical, functional properties, with the degree of similarity further shaped by learning and experience---mirroring the role $r_{1g}$ plays in the model.
 }

\subsection{Spike-and-Slab Variable Selection}
%\textcolor{magenta}{
%RGP incorporates spike-and-slab priors on the weights $\mathbf{w}^{(t)}$:
%\begin{equation}
%w_j^{(t)} \stackrel{iid}{\sim} \tilde{p} N(0,\sigma_1^2) + (1-\tilde{p})N(0,\sigma_2^2); \quad j = 1,\ldots,K; \quad t = 0,1,\ldots,T
%\label{eq:spike_slab}
%\end{equation}
%with $\tilde{p} = 0.5$, $\sigma_1^2 = 0.1$, and $\sigma_2^2 = 1$ in typical applications. This prior encourages sparsity, automatically selecting relevant inputs and intermediate representations. This provides a computational analog of attention and gain control, where the effective weight of a connection determines its contribution to prediction and error computation.}

%\textcolor{cyan}
{Layer dependence governs \emph{how} information is shared across the hierarchy; a separate question is \emph{which} inputs and intermediate representations matter at all. RGP addresses this with spike-and-slab priors on the weights $\mathbf{w}^{(t)}$:
 \begin{equation}
w_j^{(t)} \stackrel{iid}{\sim} \tilde{p} N(0,\sigma_1^2) + (1-\tilde{p})N(0,\sigma_2^2); \quad j = 1,\ldots,K; \quad t = 0,1,\ldots,T
\label{eq:spike_slab}
\end{equation}
 with $\tilde{p} = 0.5$, $\sigma_1^2 = 0.1$, and $\sigma_2^2 = 1$ in typical applications. This prior encourages sparsity, automatically selecting relevant inputs and intermediate representations, and provides a computational analog of attention and gain control: the effective weight of a connection determines its contribution to prediction and error computation, much as precision-weighting determines which prediction errors propagate through cortical hierarchies.
 }

\subsection{Bayesian Foundations and Uncertainty Quantification}
{Layer dependence and variable selection together specify RGP's architecture; realizing its Bayesian character requires propagating uncertainty through that architecture via full posterior inference. The RGP framework places priors on all parameters and performs this inference via Markov Chain Monte Carlo (MCMC). Uncertainty propagation is therefore intrinsic to the model, with posterior distributions capturing uncertainty at every hierarchy level, mirroring the brain's capacity to represent confidence; for details on the latter, see \citep{knill2004}. This Bayesian integration also confers robustness to overfitting, protecting against the overconfidence that plagues conventional neural networks \citep{bhattacharya2025}, and it supports principled model comparison via marginal likelihood---a basis for selecting among alternative hierarchical structures that is analogous to the brain's ability to choose among competing perceptual interpretations. %\Cref{fig:free-energy} illustrates how variational free energy minimization balances accuracy and complexity in this framework.
 \subsection{The Look-Up Table Approximation}
\label{subsec:lookup_table}
 Full posterior inference over a GP defined jointly on layer index and input value is computationally demanding, and RGP addresses this with a look-up table approximation. A set of auxiliary variables $\mathbf{D}_m = \{g(t_1, z_1), \ldots, g(t_m, z_m)\}$ approximates the random function $g(\cdot,\cdot)$ on a grid $\mathbf{G}_m$ \citep{bhattacharya2025}. Conditional on this approximation, the hidden units follow tractable Gaussian conditionals:
 \begin{equation}
[h_{ij}^{(t)} | \mathbf{D}_m, \mathbf{w}^{(t)}, \mathbf{h}_i^{(t-1)}, b_j^{(t)}, \boldsymbol{\theta}_g] \sim \mathcal{N}(\mu_{tij}, \sigma_{tij}^2)
\label{eq:conditional_gaussian}
\end{equation}
 This construction establishes a Markov property whereby hidden units at layer $t$ depend on lower layers only through $\mathbf{D}_m$---mirroring how higher cortical areas receive compressed representations from lower areas, rather than the full detail of lower-level activity.
 \subsection{Computational Implementation}
\label{subsec:computational_implementation}
Building on this tractable conditional structure, RGP is implemented using parallel MCMC, with full conditional distributions for key parameters derived analytically where possible to enable Gibbs sampling steps. For parameters requiring more flexible updates, the framework employs Transformation-based MCMC (TMCMC) \citep{dutta2014markov} to update fixed-dimensional parameter blocks, and Transdimensional Transformation-based MCMC (TTMCMC) \citep{das2019} to handle variable dimensionality. The parallel implementation distributes computation across multiple processors, with simulation studies demonstrating scalability up to $T = 10$ layers.
 }
%===========================================================================
% Diagram 5: RGP to Canonical Microcircuit Mapping (FINAL FIX - NO BOTTOM OVERLAPS)
% =============================================================================
%\begin{figure}[t]
\begin{figure}[htpb]
\centering
\begin{tikzpicture}[
    node distance = 1.5cm and 6cm,
    rgpcomp/.style = {rectangle, draw=blue!60, fill=blue!10, thick, minimum width=3.5cm, minimum height=0.7cm, align=center, font=\footnotesize},
    neuron/.style = {rectangle, draw=purple!60, fill=purple!20, thick, minimum width=3.5cm, minimum height=0.7cm, align=center, font=\footnotesize},
    arrow/.style = {<->, >=stealth, thick, shorten >=3pt, shorten <=3pt}
]

% Title headers
\node[align=center, font=\bfseries, text=blue] at (-3,4.5) {RGP Framework};
\node[align=center, font=\bfseries, text=purple] at (5,4.5) {Canonical Microcircuit};

% Left column: RGP Components - start higher to leave bottom space
\node[rgpcomp] (gp) at (-3,3.5) {Single GP $g(\cdot,\cdot)$};
\node[rgpcomp] (corr) at (-3,2.2) {Correlation $r_{1g}$};
\node[rgpcomp] (ss) at (-3,0.9) {Spike-and-Slab Priors};
\node[rgpcomp] (hu) at (-3,-0.4) {Hidden Units $h_{ij}^{(t)}$};
\node[rgpcomp] (like) at (-3,-1.7) {Likelihood Terms};
\node[rgpcomp] (prec) at (-3,-3.0) {Precision Parameters};
\node[rgpcomp] (mcmc) at (-3,-4.3) {MCMC Inference};
\node[rgpcomp] (lut) at (-3,-5.6) {Look-up Table $\mathbf{D}_m$};

% Right column: Neuronal Populations - spaced to match left column
\node[neuron] (dev) at (5,3.5) {Genetic/Developmental Constraints};
\node[neuron] (learn) at (5,2.2) {Experience-dependent Plasticity};
\node[neuron] (inhib) at (5,0.9) {Inhibitory Interneurons};
\node[neuron] (pyram) at (5,-0.4) {Layer 2/3 \& Layer 5 Pyramidal};
\node[neuron] (super) at (5,-1.7) {Superficial Pyramidal (L2/3)};
\node[neuron] (mod) at (5,-3.0) {Neuromodulatory Systems};
\node[neuron] (recurr) at (5,-4.3) {Recurrent Cortical Dynamics};
\node[neuron] (summ) at (5,-5.6) {Summary Statistics};

% Draw simple horizontal mapping lines
\draw[arrow] (gp) -- (dev);
\draw[arrow] (corr) -- (learn);
\draw[arrow] (ss) -- (inhib);
\draw[arrow] (hu) -- (pyram);
\draw[arrow] (like) -- (super);
\draw[arrow] (prec) -- (mod);
\draw[arrow] (mcmc) -- (recurr);
\draw[arrow] (lut) -- (summ);

% KL Divergence and Free Energy - placed in a separate box at bottom with more space
\node[draw=yellow!60, fill=yellow!10, thick, rounded corners, align=center, font=\small, text width=12cm] at (1,-7.2) {
    \textbf{KL Divergence Rate} $\mathfrak{h}(\boldsymbol{\theta}) = \log\left(\frac{\sigma_{\epsilon}}{\sigma_0}\right) - \frac{1}{2} + \frac{\sigma_0^2}{2\sigma_{\epsilon}^2} + \frac{1}{2\sigma_{\epsilon}^2}\mathbb{E}\left[(\eta_T - \eta_0)^2\right]$
};

\node[draw=yellow!60, fill=yellow!5, thick, rounded corners, align=center, font=\small, text width=12cm] at (1,-8.2) {
    \textbf{Variational Free Energy} $\displaystyle\frac{1}{n}F(q) = \mathbb{E}_q[\mathfrak{h}] + \frac{1}{n}D_{KL}[q||p] + o(1)$
};

% Connection from relevant components to KL - using curved paths to avoid overlap
\draw[->, dashed, >=stealth, thick] (hu.south) to[out=-90,in=90] (0.5,-6.8);
\draw[->, dashed, >=stealth, thick] (like.south) to[out=-90,in=90] (1.5,-6.8);
\draw[->, dashed, >=stealth, thick] (prec.south) to[out=-90,in=90] (2.5,-6.8);

% Mapping summary - moved to caption instead of figure
\end{tikzpicture}
\caption{Mapping between RGP components and neuronal populations. The shared GP corresponds to common genetic/developmental constraints across cortical areas, while $r_{1g}$ captures experience-dependent differentiation. Spike-and-slab priors map to inhibitory interneurons implementing precision-weighting. Hidden units correspond to pyramidal cell activity encoding state estimates. Likelihood terms map to superficial pyramidal cells computing prediction errors. Precision parameters correspond to neuromodulatory gain control. MCMC inference maps to recurrent cortical dynamics. The look-up table corresponds to compressed representations in higher areas. The KL divergence rate $\mathfrak{h}(\boldsymbol{\theta})$ provides a rigorous measure of precision-weighted prediction error that connects to variational free energy minimization.}
\label{fig:rgp-mapping}
\end{figure}
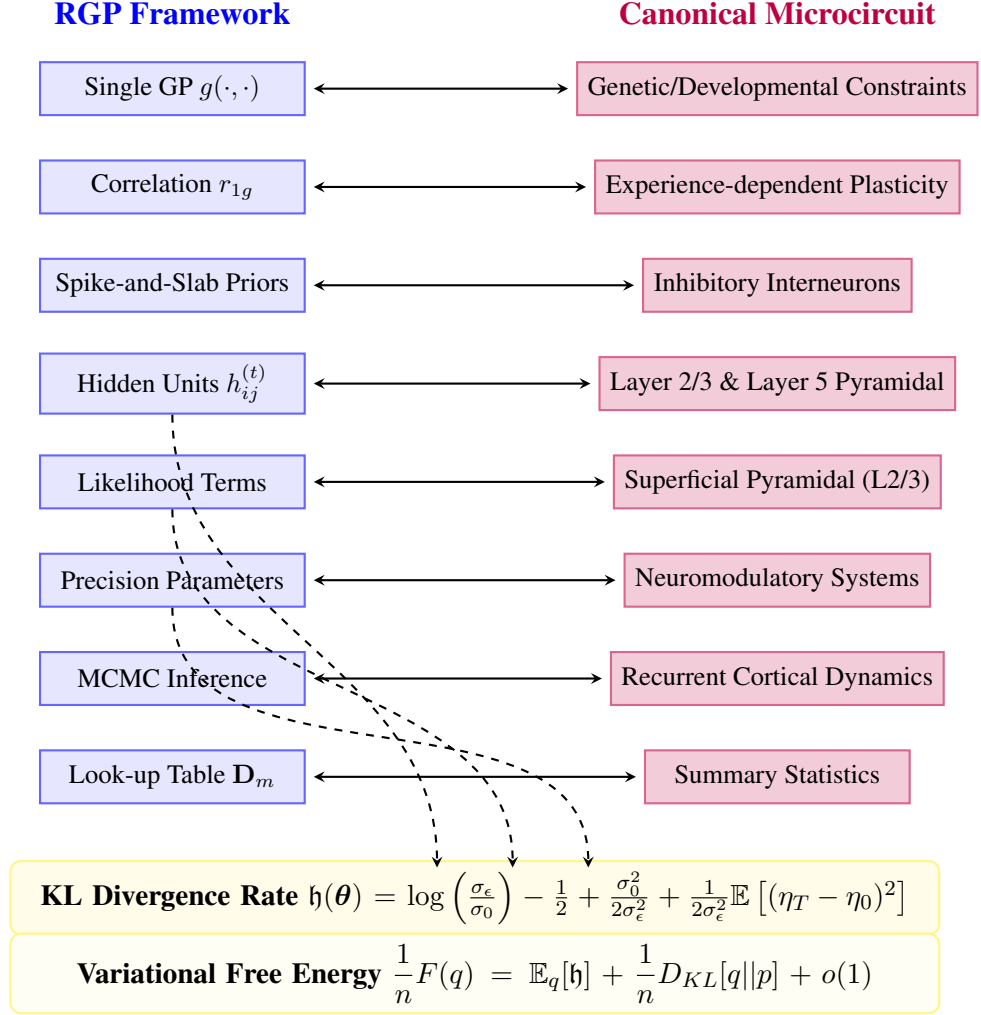

\section{Connecting RGP to Predictive Coding: A Computational Bridge}
\label{sec:connecting-rgp}

Having reviewed both the canonical microcircuit for predictive coding and the RGP framework, we now establish formal connections between them, 
%\textcolor{cyan}
{proceeding from the representational (how predictions are generated and shared across levels) to the dynamical (how errors are computed and beliefs refined over time) to the comparative (how RGP relates to other computational accounts of predictive coding). \Cref{fig:rgp-mapping} summarizes the resulting mapping.}

\subsection{Hierarchical Prediction as Recursive Function Composition with Shared Constraints}

In predictive coding, higher cortical areas generate predictions of lower-level activity through feedback connections. In RGP, higher layers generate predictions of lower-layer activity through recursive composition:

\begin{equation}
h_{ij}^{(t)} = g(t, \mathbf{w}^{(t)\prime}\mathbf{h}_i^{(t-1)} + b_j^{(t)})
\label{eq:rgp_prediction}
\end{equation}

Crucially, the functions $g(t,\cdot)$ at different layers are realizations of the same underlying GP $g(\cdot,\cdot)$, with correlation controlled by $r_{1g}$. This captures an essential biological insight: cortical areas share common architectural and developmental constraints (reflected in the shared GP) while differentiating through learning and experience (reflected in learnable $r_{1g}$).

\subsection{Bottom-Up Error Signals as Likelihood Terms}

In predictive coding, prediction errors are propagated upward to refine higher-level representations. RGP implements this through its likelihood structure. Errors are automatically weighted by precision: regions of input space where the model has high uncertainty contribute less to the overall likelihood, mirroring the brain's ability to dynamically adjust the gain of prediction error units based on reliability \citep{friston2005}.

The relationship between prediction errors and free energy minimization is evident in RGP's objective function. MCMC inference targets the posterior $p(\boldsymbol{\theta} | \mathcal{D})$, equivalent to minimizing variational free energy:

\begin{equation}
F(q) = \mathbb{E}_q[-\log p(\mathcal{D} | \boldsymbol{\theta})] + D_{KL}[q(\boldsymbol{\theta}) || p(\boldsymbol{\theta})]
\label{eq:free_energy_rgp}
\end{equation}

where the first term corresponds to expected prediction error (accuracy) and the second penalizes deviations from prior beliefs (complexity).

\subsection{Recursive Refinement Through MCMC Inference}
\label{subsec:mcmc_dynamics}

Predictive coding posits that perception involves iterative exchange of predictions and errors until the hierarchical model settles into a stable configuration \citep{friston2005}. RGP's MCMC inference implements analogous recursive refinement. At each iteration, the algorithm samples through layers, updating parameters based on information from above and below. The look-up table approximation creates a Markovian structure where each layer depends on lower layers only through sufficient statistics $\mathbf{D}_m$, mirroring how cortical areas operate on compressed representations.

A key point of discussion is the mapping between MCMC iterations and neural dynamics. While the timescales of MCMC sampling are typically longer than the millisecond-scale dynamics of neural circuits, recent theoretical work suggests that neural circuits may implement forms of sampling-based inference. \citet{buesing2011} proposed that neural populations could represent probability distributions through their stochastic activity, with the brain performing approximate Bayesian inference via sampling. The recurrent interactions in predictive coding networks have been shown to converge to posterior distributions under certain conditions \citep{millidge2020}. In this view, each MCMC iteration in RGP corresponds not to a single neural computation but to a macroscopic update of beliefs across the hierarchy, implemented by the rapid, parallel settling dynamics of the cortical circuit. The distinction between inference (updating latent states) and learning (updating parameters) in RGP mirrors the separation of fast and slow timescales observed in neural plasticity.

%\begin{comment}
%\textcolor{purple}
{A natural objection to any sampling-based account of neural inference concerns timing. Digital MCMC implementations typically require many hundreds or thousands of iterations to adequately explore a posterior, yet biological vision achieves complex, invariant object recognition within roughly 150\,ms of stimulus onset \citep{Thorpe1996}. If RGP-like sampling is to be taken seriously as a candidate account of cortical inference rather than purely as a computational tool, this apparent mismatch in timescales needs to be addressed, at least qualitatively.}
 
%\textcolor{purple}
{We do not think this objection is fatal, for four reasons, though we stress that these are qualitative arguments for why the mismatch may be smaller than it first appears, not a quantitative demonstration that cortical circuits achieve RGP-equivalent posterior convergence within 150\,ms.}
 
%\textcolor{purple}
{First, digital MCMC samplers spend a non-trivial fraction of their computation generating pseudo-random proposals. Biological neural tissue is intrinsically stochastic at multiple scales, from channel gating noise \citep{faisal2008noise} to probabilistic vesicle release \citep{branco2009probability}, which could in principle supply proposal-like variability without dedicated computation. Second, whereas a single digital MCMC chain updates sequentially, cortical populations update largely in parallel, with dense recurrent and horizontal connectivity supporting rapid information sharing across units \citep{Beck2008}; this is structurally more similar to running many interacting chains simultaneously than to a single long chain, which could substantially improve effective mixing time even without changing the per-step dynamics. Third, digital MCMC is often initialized from an arbitrary or uninformative starting point, requiring a burn-in period before samples are representative of the target distribution. Biological inference is essentially never initialized from scratch: evolutionary and developmental priors, together with a lifetime of experience-dependent plasticity, are thought to leave the system already close to a plausible region of the posterior for naturalistic stimuli \citep{pezzulo2021evolution}, which would shorten or largely eliminate the effective burn-in period rather than eliminate the need for any settling time at all. Fourth, natural sensory input is temporally continuous rather than a sequence of independent, unrelated samples; a continuous-time sampling process resembling Langevin dynamics \citep{huang2014neurons, savin2014spatio} could exploit the posterior computed at time $t-1$ as a warm start for inference at time $t$, so that each new percept requires only a local adjustment rather than exploration of the full posterior from an uninformative starting point. This picture is broadly consistent with proposals that oscillatory cortical dynamics gate discrete windows of perceptual sampling \citep{fries2015rhythms, luczak2015packet, gallina2024alpha}, though we treat the alignment between RGP's MCMC updates and specific oscillatory cycles as speculative rather than established.}
 
%\textcolor{purple}
{Together, these considerations suggest that the standard intuition about MCMC convergence times---calibrated on digital hardware with sequential, cold-started, unstructured sampling---may not transfer directly to a massively parallel, warm-started, physically stochastic substrate processing temporally correlated input. We regard this as a plausibility argument that the timescale objection is not automatically disqualifying for sampling-based accounts of fast perceptual inference, rather than as evidence that RGP's specific inference procedure achieves biologically realistic convergence speed; establishing the latter would require an explicit analysis of RGP's mixing time under parallelized, warm-started conditions, which we identify as an open question in Section~\ref{sec:discussion}.}
%\end{comment}

\subsection{Mapping RGP Components to Neuronal Populations}

\Cref{tab:mapping} summarizes the mapping between RGP components and neuronal populations.

\begin{table}[h]
\centering
\caption{Mapping between RGP components and neuronal populations}
\label{tab:mapping}
\small
\begin{tabular}{p{3.5cm}p{3.5cm}p{4cm}}
\toprule
\textbf{RGP Component} & \textbf{Neuronal Population} & \textbf{Computational Function} \\
\midrule
Single GP $g(\cdot,\cdot)$ with layer input & Genetic/developmental constraints & Shared blueprint across cortical areas \\
Correlation parameter $r_{1g}$ & Experience-dependent plasticity & Learning-dependent differentiation \\
Spike-and-slab priors on weights & Inhibitory interneurons & Implement precision-weighting \\
Hidden units $h_{ij}^{(t)} = g(t,\cdot)$ & Layer 2/3 \& layer 5 pyramidal cells & Encode current state estimates \\
Likelihood terms & Superficial pyramidal cells (L2/3) & Compute prediction errors \\
Precision parameters & Neuromodulatory systems, interneuron gain & Modulate error unit sensitivity \\
MCMC iterations & Recurrent cortical dynamics & Iterative refinement of beliefs \\
Look-up table $\mathbf{D}_m$ & Summary statistics in higher areas & Compressed representation of lower-level activity \\
\bottomrule
\end{tabular}
\end{table}

The single GP corresponds to shared developmental constraints across cortical areas. Spike-and-slab priors parallel inhibitory interneuron precision-weighting. Hidden units encode state estimates at each hierarchical level. Likelihood terms implicitly encode prediction errors, explicitly represented by superficial pyramidal cells \citep{summerfield2008}. Precision parameters correspond to gain modulation by inhibitory interneurons.

\subsection{Relation to Other Predictive Coding Implementations}
\label{subsec:pc_implementations}

The RGP framework offers a distinct but complementary perspective to other computational implementations of predictive coding. The original predictive coding model of \citet{rao1999} demonstrated how hierarchical Bayesian inference could explain extra-classical receptive field effects in visual cortex. Their model uses separate populations for predictions and errors with linear dynamics, and shows that iterative message passing leads to convergence to a solution consistent with Bayesian inference. RGP shares this emphasis on separate error and prediction units, but provides a more flexible nonparametric function for the mapping between levels via the GP.

More recently, \citet{whittington2017} showed that a predictive coding network with local Hebbian plasticity can approximate the error backpropagation algorithm. Their network implements a form of hierarchical variational inference where predictions flow downward and errors upward, with synaptic updates local to each layer. RGP's MCMC inference similarly relies on local updates---each layer's parameters are updated based only on information from adjacent layers and the shared GP representation. The key difference is that while Whittington and Bogacz's network uses fixed activation functions and learns weights through gradient descent, RGP learns both the weights and the layer-specific functions $g(t,\cdot)$ from the shared GP, providing more flexibility while maintaining locality.

These complementary perspectives suggest that predictive coding may be implemented in the brain through multiple interacting mechanisms. The RGP framework emphasizes the shared statistical structure across cortical areas and the importance of learnable similarity, while other models highlight the dynamics of inference and learning. Together, they point toward a more complete account of cortical computation.

\subsection{Addressing Degeneracy Through Learnable Layer Dependence}

RGP's solution to degeneracy has direct implications for understanding cortical hierarchies. In visual cortex, V1 extracts oriented edges, V2 combines these into more complex features, and V4/IT represent increasingly abstract object properties. Each area embodies its own specialization, with the degree of similarity between areas modifiable through learning. RGP's correlation parameter $r_{1g}$ captures this flexibility: small values allow independent representations; large values enforce similarity when beneficial.

%\subsection{\textcolor{purple}{Convergent Neurobiological Evidence for the RGP Design Choices}}
\subsection{{Convergent Neurobiological Evidence for the RGP Design Choices}}
\label{subsec:neurobio-evidence}
 
%\textcolor{purple}
{The preceding sections motivated RGP's architecture primarily on mathematical grounds: a single GP with layer-index input avoids the representational collapse that afflicts standard DGPs. Here we note that several of RGP's specific design choices are also consistent with independent empirical findings in systems and cognitive neuroscience. We present these as \emph{convergent evidence}---observations that RGP's architecture is compatible with, and in some cases naturally accommodates---rather than as confirmation that the brain implements RGP-like computations. As with the broader predictive coding mapping in Section~\ref{sec:connecting-rgp}, alternative architectures could in principle also be reconciled with each finding below; the value of the correspondences lies in their number and specificity, not in any single one being decisive.}

\begin{table}[htbp]
\centering
\caption{Qualitative comparison of RGP against machine learning and theoretical neuroscience baselines. Claims of ``advantage'' refer to architectural design choices intended to address a specific limitation, not to demonstrated empirical superiority; see the discussion following this table for caveats and open questions attached to each row.}
\label{tab:rgp_baseline_comparison}
\small
\begin{tabularx}{\textwidth}{
  >{\raggedright\arraybackslash\hsize=0.7\hsize}X 
  >{\raggedright\arraybackslash\hsize=1.1\hsize}X 
  >{\raggedright\arraybackslash\hsize=1.2\hsize}X
}
\toprule
\textbf{Competitor Class} & \textbf{Primary Limitation} & \textbf{RGP's Design Response} \\
\midrule
\multicolumn{3}{l}{\textbf{Machine Learning \& Bayesian Deep Learning Baselines}} \\
\midrule
Standard DGPs & Composing independent GP layers causes representational variance to collapse, erasing input distinctions \citep{duvenaud2014, dunlop2018}. & Uses a single shared GP $g(t,\cdot)$ with a learnable cross-layer correlation $r_{1g}$, avoiding the specific collapse mechanism that arises from independent layer composition. Whether this prevents degeneracy across all regimes of the data and hyperparameters is not established. \\
\addlinespace
Traditional DNNs & Trained via backpropagation, a procedure without strong evidence of biological implementation; produces deterministic point estimates and relies mainly on feedforward connectivity. & Performs iterative, bidirectional MCMC sampling with local conditional updates (Section~\ref{subsec:lookup_table}), structurally closer to recurrent message-passing than to feedforward point estimation. \\
\addlinespace
BNNs \& BDL \newline (e.g., MC Dropout, VI) & Rely on variational or dropout-based approximations to the posterior; typically use fixed parametric activation functions. & Performs full posterior inference via MCMC, which is asymptotically exact but---like any sampling scheme---approximate at finite sample sizes; replaces fixed activations with the learned, non-parametric function $g(t,\cdot)$. \\
\midrule
\multicolumn{3}{l}{\textbf{Theoretical Neuroscience \& Predictive Coding Baselines}} \\
\midrule
\citet{rao1999} & Implements top-down prediction via explicit subtraction, which taken literally predicts near-complete silencing of error units for well-predicted stimuli---in tension with reports of sustained, precise responses to expected stimuli \citep{spratling2008predictive, lowet2024predictive, Kok2012}. & Updates beliefs via MCMC rather than explicit subtraction. \\ %structurally compatible with sharpening-type accounts, though this does not resolve the underlying empirical debate over which mechanism dominates \citep{Kok2012}. \\
\addlinespace
\citet{whittington2017} & Uses fixed activation functions and independent scalar precision terms, which do not capture correlations between features or units. & Places a joint GP prior over $g(\cdot,\cdot)$ that induces correlation across layer index and input value (Equation~\eqref{eq:gp_cov}), and learns both weights and layer functions jointly; this is a richer correlation structure than a scalar precision.\\ %though it is not a claim of full joint covariance over all model parameters. \\
\bottomrule
\end{tabularx}
\end{table}
 
%\textcolor{purple}
{\subsubsection{High-Dimensional, Non-Degenerate Population Codes} Recall that standard DGPs suffer from a well-documented pathology: composing independent GP layers causes representational variance to collapse, erasing input distinctions. %\citep{duvenaud2014, dunlop2018}. 
Large-scale population recordings from mouse visual cortex offer a relevant empirical benchmark for what non-degenerate biological coding looks like: population responses to natural images are high-dimensional, with principal component variance decaying as a power law ($\lambda_n \propto 1/n$) rather than collapsing onto a low-dimensional subspace \citep{stringer2019high}. \citet{stringer2019high} further show mathematically that a code must maintain at least this rate of decay to remain smooth---i.e., to prevent small input perturbations from producing disproportionately large changes in population activity. This provides a concrete, quantitative notion of ``healthy" high-dimensional coding against which architectural claims about collapse can be benchmarked. RGP's use of a single shared GP with a learnable cross-layer correlation $r_{1g}$ is one architectural mechanism that avoids the pathological collapse seen in standard DGPs; whether the specific variance spectrum RGP produces across layers matches the empirical $1/n$ scaling is an open, testable question rather than a demonstrated result.}
 
%\textcolor{purple}
{\subsubsection{Continuous Macroscale Organization} RGP's covariance structure treats the layer index $t$ as an argument to a smooth kernel, $\exp\{-r_{1g}(t_1-t_2)^2/T^2\}$, which formalizes a \emph{continuously graded} notion of similarity between hierarchical levels rather than a discrete, modular one. This is qualitatively consistent with several convergent lines of evidence for continuous, rather than sharply modular, organization in cortex: a principal gradient of functional connectivity spanning from unimodal sensory cortex to transmodal association cortex \citep{Margulies2016}, a transcriptomic continuum in gene expression that tracks this same axis \citep{Burt2018}, and a corresponding gradient in intrinsic neural timescales, which lengthen progressively from sensory to association areas \citep{hasson2008, Murray2014}. Disruptions to this graded organization have also been associated with neurodevelopmental and post-lesion reorganization \citep{Huntenburg2018}. We note two caveats. First, in the current RGP formulation $t$ is evaluated only at a finite set of integer layer indices $t = 0, 1, \ldots, T$ (Equations~\eqref{eq:rgp_layer0}--\eqref{eq:rgp_output}); the GP prior itself is defined over a continuous domain, but the network architecture as implemented does not yet exploit continuously-indexed intermediate layers. The correspondence to continuous cortical gradients is therefore best read as a property of the prior RGP is built on, not (yet) a property of the fitted models in \citet{bhattacharya2025}. Second, functional connectivity gradients are macroscale, population-level phenomena measured across the whole cortex, whereas RGP's layer index is a property of a single model's internal hierarchy; the mapping is one of shared mathematical structure (smooth, graded similarity) rather than a claim that RGP layers correspond one-to-one with cortical regions along the gradient.}

\subsubsection{{Experience-Dependent Differentiation}}
RGP's covariance function separates dependence across two axes: the input-feature axis, governed by $r_{2g}$, and the cross-layer axis, governed by $r_{1g}$ (Equation~\eqref{eq:gp_cov}). This two-axis structure is loosely analogous to two developmental phenomena described in the interactive specialization literature \citep{Johnson2001}: sharpening of feature tuning with experience, and increasing functional differentiation between developing regions. Evidence for the former includes reports that infant cortical responses to faces begin broadly tuned and become more selective over development \citep{Dopierala2023}, and that multivariate response patterns to newly familiarized categories become more separable with exposure \citep{Kakaei2024}.

We emphasize that this correspondence is considerably more speculative than the two points above: it implicitly proposes a developmental trajectory for the hyperparameters $r_{1g}$ and $r_{2g}$---a trajectory that has not been directly examined in the existing RGP simulation studies. While Equation~\eqref{eq:posterior_concentration} establishes that the RGP posterior concentrates asymptotically around the truth as the sample size $n$ increases (see \citet{bhattacharya2025} for the details), this reflects statistical convergence in the limit of infinite data rather than a temporal or developmental path. Whether the intermediate stages of MCMC inference or parameter optimization correspond to biologically plausible developmental trajectories remains an open question. We present this analogy as a motivating hypothesis for future work, rather than as a validated consequence of the theory.
 
%\textcolor{purple}
{\subsubsection{Non-subtractive Error Computation} A separate consideration concerns \emph{how} prediction errors are computed, rather than how representations are organized. The original predictive coding model of \citet{rao1999} implements top-down prediction via explicit subtraction of a predicted signal from the bottom-up input. Taken literally, this predicts that well-predicted stimuli should approach complete silencing of the corresponding error units. This prediction sits awkwardly next to reports that familiar or expected stimuli can elicit sustained, precise---rather than merely attenuated---responses \citep{spratling2008predictive, lowet2024predictive}, and next to an active empirical debate over whether repetition-related response reductions reflect prediction-error suppression at all, as opposed to bottom-up adaptation or representational sharpening \citep{Kok2012}. This is a genuinely contested area of the predictive coding literature, and we do not take a position on which mechanism dominates empirically. We note it here because RGP's inference procedure does not perform explicit subtraction; predictions are updated (refined) via MCMC sampling (Section~\ref{subsec:mcmc_dynamics}), and this non-subtractive updating is at least structurally compatible with sharpening-type accounts, in which precision and reliability of tuning, rather than signal cancellation, govern the response to expected input \citep{Kok2012, Carandini2012}. Divisive normalization and marginalization-based accounts of contextual modulation offer one biophysical route by which such sharpening could be implemented without subtraction \citep{Carandini2012, beck2011marginalization}, and horizontal intra-laminar connectivity is a plausible substrate for the required contextual interactions \citep{Cohen2011}. We view this as a point of structural compatibility rather than as evidence that resolves the underlying debate.}
 
%\textcolor{purple}
{\subsubsection{Shared Uncertainty Across Populations} RGP maintains a full joint covariance structure over the single GP $g(\cdot,\cdot)$, rather than treating uncertainty at each layer or unit as independent. This design choice is consistent with the well-established observation that trial-to-trial neural response fluctuations are correlated across neurons and areas---so-called noise correlations---rather than independent \citep{Averbeck2006, Cohen2011}. We note that noise correlations are a broad, well-replicated empirical phenomenon compatible with many possible generative accounts of shared variability, not a signature unique to GP-based or hierarchical Bayesian models; we cite it here as a minimal consistency check (a model with strictly independent layer-wise or unit-wise uncertainty would be harder to reconcile with this literature) rather than as strong confirmatory evidence for RGP specifically. Theoretical work on sampling-based codes in recurrent circuits \citep{buesing2011, orban2016} offers one account of how such correlated variability could arise from, and support, approximate Bayesian inference of the kind RGP's MCMC implements.}

\section{The Free Energy Principle and RGP Inference: A Rigorous Foundation}
\label{sec:free-energy}

%The free energy principle \citep{friston2010} provides a unifying theoretical framework connecting predictive coding to fundamental physical principles. Here we show how RGP inference can be rigorously understood within this framework.

%\textcolor{cyan}
{The free energy principle \citep{friston2010} provides a unifying theoretical framework connecting predictive coding to fundamental physical principles. It is a central claim of the free energy principle that any self-organizing system maintaining integrity over time must minimize the surprise of its sensory states, and that this quantity is upper-bounded by variational free energy. Here we show how RGP inference can be situated within this framework, and we make explicit which parts of the resulting picture are established results and which are heuristic correspondences.
 }

\subsection{Notation and Setup}
\label{subsec:notation}

Before proceeding, we establish the necessary notation. Let $\boldsymbol{\theta}$ denote the collection of all parameters in the RGP model, including the GP hyperparameters, weights, biases, and precision parameters. Let $\boldsymbol{\theta}_0$ denote the true (but unknown) parameter values that generate the observed data. The true data-generating distribution is $p_{\theta_0}(\mathbf{y}_n)$, while our model-implied distribution is $p_{\theta}(\mathbf{y}_n)$. %\textcolor{cyan}
{We write $Q$ for the dimension of the output $\mathbf{y}$ (i.e., $\mathbf{y}_n \in \mathbb{R}^Q$); this should not be confused with the variational density $q(\boldsymbol\theta)$ introduced below, whose lowercase symbol and explicit argument make its role unambiguous throughout.}

For a given input $\mathbf{x}$, define the model prediction after $T$ layers of recursive composition as $\eta_T(\mathbf{x}, \dot{b}_j)$, where $\dot{b}_j$ are the output bias parameters for output coordinate $j = 1,\ldots,Q$ (see Equations \eqref{eq:rgp_layer0}--\eqref{eq:rgp_output}). The true regression function is denoted $\eta_0(\mathbf{x}, c_j)$, where $c_j$ represents the true bias for coordinate $j$. The observation noise variance in the model is $\sigma_{\epsilon}^2$, while $\sigma_0^2$ denotes the true observation noise variance. The expectation $\mathbb{E}_{\mathbf{X}}$ is taken with respect to the distribution of inputs (either empirical or true).

Let $\boldsymbol\Theta$ denote the full parameter space. For any subset $\boldsymbol A \subseteq \boldsymbol\Theta$, define the essential infimum of the KL divergence rate over $A$ as:
\begin{equation}
\mathfrak{h}(\boldsymbol A) = \underset{\boldsymbol\theta \in \boldsymbol A}{\mathrm{essinf}} \, \mathfrak{h}(\boldsymbol\theta)
\label{eq:essinf}
\end{equation}
In particular, $\mathfrak{h}(\boldsymbol\Theta)$ is the minimum achievable KL divergence rate over the entire parameter space; the results below show that this quantity plays the role of an irreducible prediction-error floor for the RGP model.

\subsection{Variational Free Energy and KL Divergence}

Variational free energy is defined as

\begin{equation}
F(q) = \mathbb{E}_{q(\boldsymbol{\theta})}[\log q(\boldsymbol{\theta}) - \log p(\mathbf{y}, \boldsymbol{\theta})]
\label{eq:free_energy_definition}
\end{equation}

where $q(\boldsymbol{\theta})$ approximates the posterior, and $p(\mathbf{y}, \boldsymbol{\theta})$ is the generative model. To connect this to prediction error, we first need the asymptotic behavior of the model likelihood, which is governed by the KL divergence rate $\mathfrak{h}(\boldsymbol{\theta})$ \citep{bhattacharya2025}:

\begin{equation}
\mathfrak{h}(\boldsymbol{\theta}) = \lim_{n\to\infty} \frac{1}{n} \mathbb{E}_{\boldsymbol{\theta}_0}\left[\log \frac{p_{\boldsymbol{\theta}_0}(\mathbf{y}_n)}{p_{\boldsymbol{\theta}}(\mathbf{y}_n)}\right]
\label{eq:kl_rate_definition}
\end{equation}

For the RGP model, this rate can be computed explicitly \citep[Equation (90)]{bhattacharya2025}:

\begin{equation}
\mathfrak{h}(\boldsymbol{\theta}) = \log\left(\frac{\sigma_{\epsilon}}{\sigma_0}\right) - \frac{1}{2} + \frac{\sigma_0^2}{2\sigma_{\epsilon}^2} + \frac{1}{2\sigma_{\epsilon}^2}\sum_{j=1}^Q \mathbb{E}_{\mathbf{X}}\left[\left(\eta_T(\mathbf{X},\dot{b}_j) - \eta_0(\mathbf{X},c_j)\right)^2\right]
\label{eq:rgp_kl_rate_explicit}
\end{equation}

%\textcolor{cyan}
{The first three terms depend only on $\sigma_\epsilon$ and $\sigma_0$, and vanish exactly when $\sigma_\epsilon = \sigma_0$; they measure how well the model's noise level matches the true noise level. The final term is the squared prediction error, summed over the $Q$ output coordinates and averaged over the input distribution, weighted by the precision $1/(2\sigma_{\epsilon}^2)$. Read this way, \eqref{eq:rgp_kl_rate_explicit} decomposes the KL rate into a noise-calibration cost and a precision-weighted prediction-error cost — the same two ingredients that appear in the predictive coding formulation below.}

\subsection{From KL Divergence Rate to Variational Free Energy}

\citet{shalizi2009} and, for the RGP model specifically, \citet{bhattacharya2025} establish that the normalized log likelihood ratio converges almost surely:

\begin{equation}
\frac{1}{n}\log R_n(\boldsymbol{\theta}) = \frac{1}{n}\log\frac{p_{\boldsymbol{\theta}}(\mathbf{y}_n)}{p_{\boldsymbol{\theta}_0}(\mathbf{y}_n)} \xrightarrow{a.s.} -\mathfrak{h}(\boldsymbol{\theta})
\label{eq:llr_convergence}
\end{equation}
Substituting $\log p(\mathbf{y},\boldsymbol\theta) = \log p(\boldsymbol\theta) + \log p_{\boldsymbol\theta}(\mathbf y_n)$ into \eqref{eq:free_energy_definition} and applying \eqref{eq:llr_convergence} term by term gives:

\begin{equation}
\boxed{\frac{1}{n}F(q) = \underbrace{\mathbb{E}_q[\mathfrak{h}(\boldsymbol{\theta})]}_{\text{Expected Prediction Error}} + \underbrace{\frac{1}{n}D_{KL}[q(\boldsymbol{\theta}) \,\|\, p(\boldsymbol{\theta})]}_{\text{Complexity Penalty}} + \, C_n + o(1)}
\label{eq:free_energy_decomposition}
\end{equation}

where $C_n = -\frac{1}{n}\log p_{\boldsymbol\theta_0}(\mathbf y_n)$ is the entropy rate of the true data-generating process. Because $C_n$ does not depend on $\boldsymbol\theta$ or on the variational family $q(\boldsymbol\theta)$, it plays no role in the optimization over $q$: minimizing $F(q)$ is equivalent to minimizing the expected KL divergence rate (prediction error) plus the complexity penalty. 

\subsection{Precision-Weighting and the Geometry of Prediction Error}

The precision weighting made explicit in \eqref{eq:rgp_kl_rate_explicit} — prediction error scaled by $1/(2\sigma_{\epsilon}^2)$ — mirrors the precision-weighted prediction error central to predictive coding \citep{friston2009}: predictions are penalized in proportion to how confidently (i.e., with what precision) the model expects to have gotten them right.

This correspondence has a consequence for how the posterior behaves. Define the neighborhood of parameters with KL divergence rate within $\epsilon_n$ of the infimum:

\begin{equation}
N_{\epsilon_n} = \{\boldsymbol{\theta} : \mathfrak{h}(\boldsymbol{\theta}) \leq \mathfrak{h}(\boldsymbol{\Theta}) + \epsilon_n\}
\label{eq:neighborhood}
\end{equation}

with $\epsilon_n \to 0$ and $n\epsilon_n \to \infty$ as $n \to \infty$. Under these conditions, together with the prior mass and entropy conditions of \citet{bhattacharya2025}, the posterior concentrates on this neighborhood almost surely:

\begin{equation}
\lim_{n\to\infty} \pi(N_{\epsilon_n} \mid \mathbf{y}_{nQ}) = 1
\label{eq:posterior_concentration}
\end{equation}

In free-energy terms: as more data arrive, the posterior increasingly rules out parameter values whose expected prediction error exceeds the achievable floor $\mathfrak h(\boldsymbol\Theta)$, which is exactly the behavior the decomposition in \eqref{eq:free_energy_decomposition} would lead us to expect.

\subsection{Robustness to Model Misspecification}

We now ask how small the irreducible floor $\mathfrak h(\boldsymbol\Theta)$ can be made, including in cases where the true generative model does not belong to the assumed function class. The universal approximation theorem guarantees the existence of approximating functions: for any $\epsilon > 0$, there exist functions $\tilde{f}(\cdot)$ and $\tilde{g}(t,\cdot)$ for layer index $t = 0,\ldots,T$ and parameters $\{\tilde{w}_j^{(t)}, \tilde{b}_j^{(t)}\}$ such that the resulting hierarchical composition $\tilde{\eta}_T$ satisfies:

\begin{equation}
\|\tilde{\eta}_T(\cdot, c_j) - \eta_0(\cdot, c_j)\| < \epsilon
\label{eq:universal_approximation}
\end{equation}

for sufficiently large recursion depth $T$, where $\tilde{\eta}_T$ is obtained by composing $\tilde f$ and $\tilde g$ according to the RGP architecture. Because the prediction-error term in \eqref{eq:rgp_kl_rate_explicit} is a function of exactly this approximation error, \eqref{eq:universal_approximation} implies that $\mathfrak{h}(\boldsymbol{\Theta})$ can be driven arbitrarily close to zero by increasing $T$. We use this bound once more below and do not restate it.

To make the resulting predictive guarantee precise, we define the true one-step-ahead predictive distribution $P^{nQ}$ as the conditional distribution of the next observation given the past under the true data-generating mechanism:

\begin{equation}
P^{nQ} = P\left(\{y_{nj}; j = 1,\ldots,Q\} \mid \{y_{1j},\ldots,y_{n-1,j}; j = 1,\ldots,Q\}\right)
\label{eq:predictive_true}
\end{equation}

and the posterior predictive distribution $F_{\pi}^{nQ}$, which averages the model's predictive distributions over the posterior:

\begin{equation}
F_{\pi}^{nQ} = \int_{\Theta} F_{\theta}^{nQ} \, d\pi(\boldsymbol{\theta} \mid \mathbf{y}_{nQ})
\label{eq:predictive_posterior}
\end{equation}

where $F_{\theta}^{nQ}$ is the one-step-ahead predictive distribution under parameter $\theta$. We measure the discrepancy between $P^{nQ}$ and $F_\pi^{nQ}$ using the squared Hellinger distance, defined generically for two probability measures $\mu$ and $\nu$ as

\begin{equation}
\rho_H^2(\mu,\nu) = \frac{1}{2} \int \left(\sqrt{d\mu} - \sqrt{d\nu}\right)^2 \in [0,1],
\label{eq:hellinger}
\end{equation}

with $\rho_H^2 = 0$ indicating identical distributions and $\rho_H^2 = 1$ indicating disjoint support.

Under assumptions (A1)--(A5), building on \citet{shalizi2009} and established for the RGP model in \citet[Theorem 4]{bhattacharya2025}, the limit superior of the posterior predictive discrepancy is bounded almost surely by the KL divergence floor, for both Hellinger and total variation distance:

\begin{equation}
\limsup_{n\to\infty} \rho_H^2\!\left(P^{nQ}, F_{\pi}^{nQ}\right) \leq \mathfrak{h}(\boldsymbol{\Theta}) < \epsilon, \qquad
\limsup_{n\to\infty} \rho_{TV}^2\!\left(P^{nQ}, F_{\pi}^{nQ}\right) \leq 4\,\mathfrak{h}(\boldsymbol{\Theta}) < 4\epsilon
\quad \text{a.s.}
\label{eq:predictive_convergence}
\end{equation}

%\textcolor{cyan}
{
Combined with \eqref{eq:universal_approximation}, this shows that increasing recursion depth $T$ drives both upper bounds toward zero, so that the RGP model's posterior predictive distribution tracks the true one-step-ahead predictive distribution arbitrarily closely, in the long run, even when the true generative process lies outside the assumed function class.

The preceding asymptotic results establish that, as data accumulate, RGP posterior inference behaves in a manner formally consistent with free-energy minimization: expected prediction error is driven toward an irreducible floor, and that floor can be made arbitrarily small by increasing recursion depth. This raises a further question, addressed next: what does this formal correspondence license us to claim about the brain itself?
}

% =============================================================================
% Diagram 6: Free Energy Minimization and KL Divergence Rate (NO OVERLAPS - GUARANTEED)
% =============================================================================
%\begin{figure}[t]
\begin{figure}[htpb]
\centering
\begin{tikzpicture}

% Main vertical column - spaced far apart
% Top box at y=5
\node[draw, thick, fill=blue!10, rounded corners, minimum width=8cm, minimum height=1.5cm, align=center] (top) at (0,5) {
    \textbf{Generative Model} $p(\mathbf{y},\boldsymbol{\theta})$ \hspace{1cm} \textbf{Posterior} $q(\boldsymbol{\theta})$
};

% KL box at y=1 (4 units below top)
\node[draw, thick, fill=yellow!20, rounded corners, minimum width=8cm, minimum height=2cm, align=center] (kl) at (0,1) {
    \textbf{KL Divergence Rate} $\mathfrak{h}(\boldsymbol{\theta})$\\[4pt]
    $\mathfrak{h}(\boldsymbol{\theta}) = \log\left(\frac{\sigma_{\epsilon}}{\sigma_0}\right) - \frac{1}{2} + \frac{\sigma_0^2}{2\sigma_{\epsilon}^2} + \frac{1}{2\sigma_{\epsilon}^2}\sum_{j=1}^q \mathbb{E}_{\mathbf{X}}\left[(\eta_T - \eta_0)^2\right]$
};

% Asymptotic box at y=-3 (4 units below KL)
\node[draw, thick, fill=green!20, rounded corners, minimum width=8cm, minimum height=1.5cm, align=center] (asymp) at (0,-3) {
    \textbf{Asymptotic Relation}\\[4pt]
    $\frac{1}{n}F(q) = \mathbb{E}_q[\mathfrak{h}] + \frac{1}{n}D_{KL}[q||p] + o(1)$
};

% Main vertical arrows - using coordinates in the empty spaces
\draw[->, thick] (0,4) -- (0,2.5);
\draw[->, thick] (0,0) -- (0,-1.5);

% Left side box - far left at x=-7
\node[draw, thick, fill=red!20, rounded corners, minimum width=3.5cm, minimum height=1.2cm, align=center] (prec) at (-7,1) {
    \textbf{Precision}\\
    $\frac{1}{2\sigma_{\epsilon}^2}$
};

% Right side boxes - far right at x=7, stacked with vertical gap
\node[draw, thick, fill=purple!20, rounded corners, minimum width=3.5cm, minimum height=1.2cm, align=center] (fe) at (7,2.5) {
    \textbf{Free Energy}\\
    $F(q)$
};

\node[draw, thick, fill=cyan!20, rounded corners, minimum width=3.5cm, minimum height=1.2cm, align=center] (conc) at (7,-0.5) {
    \textbf{Concentration}\\
    $\pi(N_{\epsilon_n}|\mathbf{y}) \to 1$
};

% Connections from sides to KL - using clear horizontal then vertical paths
% Left to KL
\draw[->, thick] (-5.5,1) -- (-3,1) -- (-3,1) -- (-1,1);

% Right to KL - two separate paths at different y-levels
\draw[->, thick] (5.5,2.5) -- (3,2.5) -- (3,1.3);
\draw[->, thick] (5.5,-0.5) -- (3,-0.5) -- (3,0.7);

% LLN at bottom - far below everything
\node[align=center, font=\small] at (0,-5) {
    $\frac{1}{n}\log R_n(\boldsymbol{\theta}) \xrightarrow{a.s.} -\mathfrak{h}(\boldsymbol{\theta})$
};

% Add a light grid to verify no overlaps (optional, can be removed)
% \draw[step=1cm, gray!10, very thin] (-8,-6) grid (8,6);

\end{tikzpicture}
\caption{Free energy minimization in RGP, showing the connection to the KL divergence rate $\mathfrak{h}(\boldsymbol{\theta})$. RGP inference minimizes variational free energy through MCMC sampling.}
\label{fig:free-energy}
\end{figure}
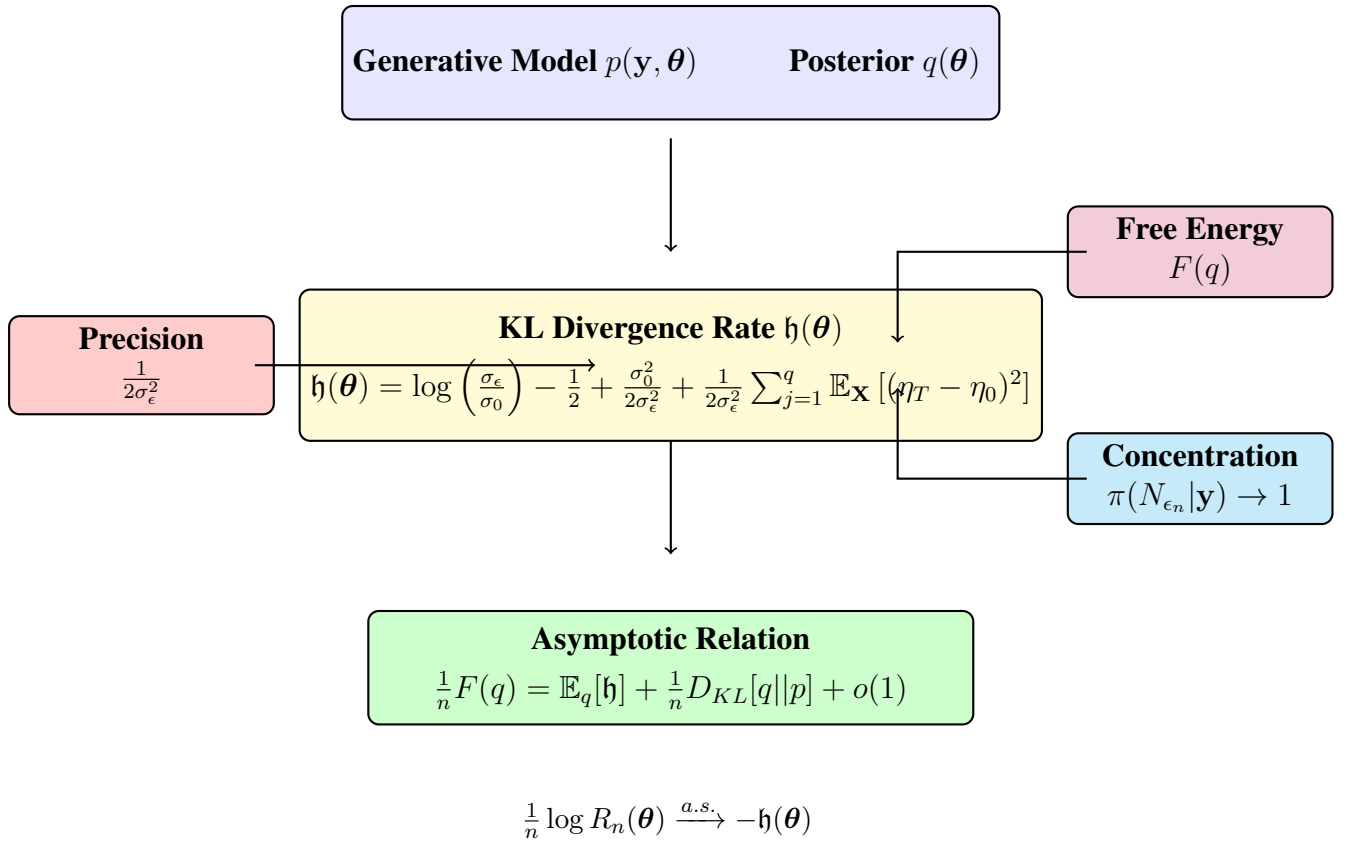

\section{Epistemological Status of Bayesian Models}
\label{sec:epistemological}

\citet{colombo2012} provide a penetrating analysis of the epistemological status of Bayesian models in neuroscience, distinguishing between instrumentalist and realist interpretations.

\subsection{Instrumentalist vs. Realist Interpretations}

Instrumentalism holds that scientific models are useful instruments for predicting and systematizing observations, but need not correspond to actual mechanisms. A Bayesian model of perception is valuable if it accurately predicts behavior, regardless of whether the brain actually implements Bayesian computations.

Realism holds that successful models correspond to actual mechanisms. A realist interpretation would claim the brain genuinely represents probability distributions and performs Bayesian inference.

\citet{colombo2012} argue that current psychophysical evidence supports only an instrumentalist interpretation. Studies showing optimal Bayesian behavior \citep{ernst2002, knill2004} demonstrate that Bayesian models predict behavior, but do not establish that the brain implements Bayesian computations. Alternative non-Bayesian algorithms can produce identical behavior \citep{maloney2009}.

\subsection{From Behavior to Mechanisms}

Moving from instrumentalist to realist interpretation requires linking behavioral-level models to mechanistic-level evidence. The RGP framework, combined with canonical microcircuit mapping, contributes to this project. By providing a mathematically rigorous specification of hierarchical Bayesian inference, and showing how its components map onto identified neuronal populations, RGPs help bridge algorithmic and implementational levels. The single GP with layer-index input corresponds to shared developmental constraints; $r_{1g}$ to experience-dependent differentiation; spike-and-slab priors to precision-weighting by inhibitory interneurons; MCMC dynamics to recurrent cortical interactions. \Cref{fig:rgp-mapping} summarizes this mapping.

\section{Simulation Results and Empirical Phenomena}
\label{sec:simulation-results}

The original RGP paper \citep{bhattacharya2025} presents comprehensive simulation studies across architectures with different depths ($T = 1$ to $10$), providing rich insights into how hierarchical Bayesian inference manifests in practice.

\subsection{Predictive Performance and Hierarchical Depth}

Figure 5 in the original RGP paper shows posterior predictions for held-out data across network depths. Best predictive performance is achieved for $T = 4$, with $T = 1$ also performing well, while deeper architectures ($T \geq 6$) show deteriorating performance. This pattern aligns with predictive coding: each hierarchical level refines predictions by minimizing local prediction errors, with optimal depth representing the point where additional layers provide diminishing returns. The $T = 4$ optimum suggests sufficient depth to propagate prediction errors upward and refine predictions downward through enough iterations to achieve convergence to the posterior.

\subsection{The Role of the Number of Basis Functions $L$}

Figures 7-8 of \citep{bhattacharya2025} display trace plots of $L$, the number of basis functions in the Karhunen-Loève expansion of the output GP. Best predictive models ($T = 1,4$) exhibit posterior concentration of $L$ at relatively large values ($L = 8,10$), while poorer models show high variability or concentration at small values ($L = 2$ for $T = 5$). This finding has a direct analog in predictive coding's precision-weighting mechanism: $L$ controls the complexity of output function representation, corresponding to the precision with which predictions can be generated. When $L$ is too small, the model lacks representational capacity; when $L$ is highly variable, the model fails to settle on a stable representation.

\subsection{Repetition Suppression and Mismatch Negativity}

Repetition suppression---reduction in neural responses to repeated stimuli---depends on expectations, with predictable repetitions producing greater suppression \citep{summerfield2008}. In RGP, consider MCMC sampling for a fixed input presented repeatedly. At first presentation, posterior over hidden units and parameters exhibits high uncertainty; prediction errors are large. As the stimulus repeats, predictions become increasingly accurate, and MCMC updates produce smaller changes as the posterior converges. The rate of convergence depends on prediction precision, matching empirical findings.

Mismatch negativity---enhanced response to deviant stimuli---occurs when a stimulus violates expectations. In RGP, after repeated presentations of a regular stimulus, posterior concentrates around parameters minimizing prediction error for that stimulus. When a deviant stimulus appears, likelihood under those parameters is low, generating large prediction error, corresponding to enhanced neural response.

\subsection{Laminar-Specific Dynamics and Spectral Asymmetries}
\label{subsec:laminar_dynamics}

The correlation parameter $r_{1g}$ provides an additional prediction: the degree of functional similarity between layers should be learnable and task-dependent. When a task requires independent processing at different levels, $r_{1g}$ should decrease; when tight coupling is needed, $r_{1g}$ should increase. This predicts that functional connectivity between cortical areas should be dynamically modulated by task demands, observed empirically \citep{buchel1999, friston1997}.

% =============================================================================
% Diagram 7: Traditional Deep Learning vs RGP vs Predictive Coding (FIXED - NO CURTAILED TEXT)
% =============================================================================
%\begin{figure}[t]
\begin{figure}[htpb]
\centering
\begin{tikzpicture}[
    node distance = 1.2cm and 2.5cm,
    header/.style = {rectangle, draw=black!80, fill=gray!20, thick, minimum width=4cm, minimum height=1.2cm, align=center, font=\bfseries\small},
    box/.style = {rectangle, draw=black!60, thick, minimum width=4cm, minimum height=1cm, align=center, font=\small},
    rowlabel/.style = {align=right, font=\small\bfseries, text width=3.5cm},
    arrow/.style = {->, >=stealth, thick, shorten >=2pt, shorten <=2pt}
]

% Column headers at the top
\node[header] (tdl) at (-2,3) {\textbf{Traditional Deep Learning}};
\node[header] (rgp) at (3,3) {\textbf{Recursive GP}};
\node[header] (pc) at (8,3) {\textbf{Predictive Coding}};

% Row labels on left - with wider text width and positioned further left
\node[rowlabel] at (-6.5,2) {Information flow:};
\node[rowlabel] at (-6.5,1) {Representation:};
\node[rowlabel] at (-6.5,0) {Layer dependence:};
\node[rowlabel] at (-6.5,-1) {Learning:};
\node[rowlabel] at (-6.5,-2) {Error handling:};
\node[rowlabel] at (-6.5,-3) {Uncertainty:};
\node[rowlabel] at (-6.5,-4) {Attention:};
\node[rowlabel] at (-6.5,-5) {Biological plausibility:};

% Traditional Deep Learning column entries
\node[box, fill=gray!5] (tdl1) at (-2,2) {Feedforward};
\node[box, fill=gray!5] (tdl2) at (-2,1) {Deterministic};
\node[box, fill=gray!5] (tdl3) at (-2,0) {Fixed composition};
\node[box, fill=gray!5] (tdl4) at (-2,-1) {Backpropagation (global)};
\node[box, fill=gray!5] (tdl5) at (-2,-2) {Uniform weighting};
\node[box, fill=gray!5] (tdl6) at (-2,-3) {Not intrinsic};
\node[box, fill=gray!5] (tdl7) at (-2,-4) {External mechanism};
\node[box, fill=gray!5] (tdl8) at (-2,-5) {Low};

% RGP column entries
\node[box, fill=blue!5] (rgp1) at (3,2) {Recurrent message passing};
\node[box, fill=blue!5] (rgp2) at (3,1) {Single GP $g(\cdot,\cdot)$};
\node[box, fill=blue!5] (rgp3) at (3,0) {Learnable via $r_{1g}$};
\node[box, fill=blue!5] (rgp4) at (3,-1) {Local MCMC updates};
\node[box, fill=blue!5] (rgp5) at (3,-2) {Precision-weighted};
\node[box, fill=blue!5] (rgp6) at (3,-3) {Full posterior};
\node[box, fill=blue!5] (rgp7) at (3,-4) {Via spike-and-slab};
\node[box, fill=blue!5] (rgp8) at (3,-5) {High};

% Predictive Coding column entries
\node[box, fill=green!5] (pc1) at (8,2) {Bidirectional};
\node[box, fill=green!5] (pc2) at (8,1) {Probabilistic};
\node[box, fill=green!5] (pc3) at (8,0) {Experience-dependent};
\node[box, fill=green!5] (pc4) at (8,-1) {Local Hebbian};
\node[box, fill=green!5] (pc5) at (8,-2) {Precision-weighted};
\node[box, fill=green!5] (pc6) at (8,-3) {Essential};
\node[box, fill=green!5] (pc7) at (8,-4) {Gain modulation};
\node[box, fill=green!5] (pc8) at (8,-5) {High};

% Status labels at the bottom
\node[align=center, font=\small\itshape, text=red] at (-2,-6.2) {Does NOT implement\\predictive coding};
\node[align=center, font=\small\itshape, text=blue] at (3,-6.2) {Implements\\predictive coding};
\node[align=center, font=\small\itshape, text=green!50!black] at (8,-6.2) {Canonical\\microcircuit};

% Connection arrows between columns
\draw[arrow, red!50, thick] (0,-6.2) -- node[above, sloped, font=\tiny] {Poor match} (6.5,-6.2);
\draw[arrow, green!70!black, thick] (4.5,-6.2) -- node[above, sloped, font=\tiny] {Strong correspondence} (7,-6.2);
\draw[arrow, blue!50, thick] (0,-6.8) -- node[below, sloped, font=\tiny] {Inspiration} (4.5,-6.8);

% Add a background grid for better readability
\foreach \y in {2,1,0,-1,-2,-3,-4,-5} {
    \draw[gray!20, thick] (-7.5,\y+0.3) -- (10,\y+0.3);
}

% Add column separators
\draw[dashed, gray!50] (0.5,3.5) -- (0.5,-6);
\draw[dashed, gray!50] (5.5,3.5) -- (5.5,-6);

% Add left boundary guide (optional - shows nothing is cut off)
\draw[gray!10] (-7.5,4) -- (-7.5,-7);

\end{tikzpicture}
\caption{Comparison of traditional deep learning, RGP, and predictive coding principles across key computational dimensions. Traditional deep learning diverges from predictive coding in its feedforward architecture, deterministic representations, global learning rules, and uniform error weighting. RGP aligns closely through recurrent message passing, probabilistic representations with a single shared GP, local MCMC updates mirroring synaptic plasticity, and precision-weighted error handling via spike-and-slab priors. The learnable correlation parameter $r_{1g}$ captures experience-dependent differentiation between layers, directly mapping to the canonical microcircuit's proposal that cortical areas share developmental constraints while specializing through learning.}
\label{fig:comparison}
\end{figure}

\section{Comparison of RGP with Traditional Deep Learning Models and Predictive Coding}
\label{sec:comparison-traditional}

\Cref{fig:comparison} provides a visual summary of the comparison between traditional deep learning, RGP, and predictive coding.

\subsection{Architectural Differences}

Traditional deep learning models are organized as feedforward hierarchies, with information flowing from input to output through successive layers. This contrasts with predictive coding's recurrent message passing between hierarchical levels. RGP embodies this recurrent architecture: MCMC inference involves iterative sampling through the hierarchy, with each pass updating representations based on information from both higher and lower levels.

\subsection{Probabilistic Foundations}

Standard deep networks are deterministic function approximators outputting point estimates. While techniques like dropout \citep{gal2016} can be interpreted as approximate Bayesian inference, these are post-hoc additions. RGP is fundamentally probabilistic: the single GP maintains a full Gaussian process prior over functions at all layers, enabling principled uncertainty propagation.

\subsection{Learning Mechanisms}

Traditional deep networks are trained via backpropagation, requiring global error signals and non-local learning rules, lacking biological plausibility \citep{crick1989}. RGP employs local MCMC updates depending only on information from neighboring layers, aligning with synaptic plasticity principles. %\citep{dayan2005}.

\subsection{Precision-Weighting and Attention}

Predictive coding features precision-weighting---modulation of prediction error gain based on reliability \citep{feldman2010}. Traditional deep networks have no intrinsic precision-weighting mechanism. RGP naturally incorporates precision-weighting through the covariance structure of its GP priors and spike-and-slab variable selection.

\subsection{Correspondence with Predictive Coding Principles}

\Cref{tab:comparison} summarizes the correspondence between traditional deep learning, RGP, and predictive coding.

\begin{table}[h]
\centering
\caption{Comparison of traditional deep learning, RGP, and predictive coding principles}
\label{tab:comparison}
\footnotesize
\begin{tabular}{p{2.5cm}p{2.5cm}p{2.5cm}p{2.5cm}}
\toprule
\textbf{Principle} & \textbf{Traditional DL} & \textbf{RGP Framework} & \textbf{Predictive Coding} \\
\midrule
Information flow & Primarily feedforward & Recurrent message passing & Bidirectional \\
Representation & Deterministic point estimates & Single GP $g(\cdot,\cdot)$ with layer input & Probabilistic with shared constraints \\
Layer dependence & Fixed composition & Learnable via $r_{1g}$ & Experience-dependent \\
Learning & Backpropagation (global) & Local MCMC updates & Local Hebbian plasticity \\
Error handling & Uniform error weighting & Precision-weighted & Precision-weighted \\
Uncertainty & Not intrinsic & Intrinsic; full posterior & Essential for optimal inference \\
Attention & External mechanism & Precision-weighting via spike-and-slab & Gain modulation by inhibitory interneurons \\
Biological plausibility & Low & High & High \\
\bottomrule
\end{tabular}
\end{table}

\section{Bayesian Deep Learning, Bayesian Neural Networks, and RGP}
\label{sec:comparison-bayesian}

\subsection{Bayesian Neural Networks: Foundations and Limitations}

Bayesian Neural Networks (BNNs) place prior distributions over weights, performing Bayesian inference to obtain posteriors \citep{neal1996, macKay1992}. Predictions integrate over the posterior:

\begin{equation}
p(y^*|\mathbf{x}^*,\mathcal{D}) = \int p(y^*|\mathbf{x}^*,\mathbf{w}) p(\mathbf{w}|\mathcal{D}) d\mathbf{w}
\end{equation}

BNNs provide uncertainty estimates and naturally regularize, but face limitations: computational intractability, %lack of hierarchical structure, 
fixed activation functions, and difficulty propagating uncertainty through layers.

\subsection{Bayesian Deep Learning: Variational Methods}

Bayesian deep learning encompasses techniques like Monte Carlo dropout \citep{gal2016} and variational inference for BNNs \citep{graves2011, blundell2015}. While scalable, these provide approximations that may underestimate uncertainty and lack biological constraints.

\subsection{Relevance to Predictive Coding}

Points of contact include uncertainty quantification, hierarchical structure, and free energy minimization. However, standard Bayesian deep learning lacks predictive coding's specific architectural constraints: separate populations for predictions and errors, laminar-specific connectivity, precision-weighting mechanisms, and local learning rules.

\Cref{tab:bayesian-comparison} provides a systematic comparison.

\begin{table}[h]
\centering
\caption{Comparison of Bayesian neural networks, Bayesian deep learning, RGP, and predictive coding}
\label{tab:bayesian-comparison}
\footnotesize
\begin{tabular}{p{2.2cm}p{2.2cm}p{2.2cm}p{2.2cm}p{2.2cm}}
\toprule
\textbf{Principle} & \textbf{BNNs} & \textbf{Bayesian Deep Learning} & \textbf{RGP Framework} & \textbf{Predictive Coding} \\
\midrule
Uncertainty representation & Posterior over weights & Approximate posterior & Single GP $g(\cdot,\cdot)$ with layer input & Precision-weighted distributions \\
Layer dependence & Independent & Independent & Learnable via $r_{1g}$ & Experience-dependent \\
Hierarchical structure & Implicit via depth & Implicit via depth & Recursive with shared GP & Area-specific with common constraints \\
Inference & Intractable; MCMC/VI & Variational approximations & Parallel MCMC & Recurrent message passing \\
Learning & Global (backprop) & Global (backprop/VI) & Local MCMC updates & Local Hebbian plasticity \\
Error handling & Through likelihood & Through likelihood & Precision-weighted & Precision-weighted \\
Activation functions & Fixed (ReLU, tanh) & Fixed & Learnable GP functions $g(t,\cdot)$ & Learnable neural responses \\
Precision-weighting & Not present & Not intrinsic & Via spike-and-slab & Via inhibitory interneurons \\
Biological plausibility & Low & Low & High & High \\
Scalability & Limited & Good & Moderate & N/A (theoretical) \\
\bottomrule
\end{tabular}
\end{table}

\section{Parallel Computation in RGPs and Neurobiological Processing}
\label{sec:parallel-computation}

\subsection{Parallel MCMC as a Model of Neural Dynamics}

RGP employs parallel MCMC, where computations for each iteration are distributed across multiple processors working together on a single chain. This mirrors the brain's distributed processing architecture: different cortical columns and areas operate concurrently on different aspects of sensory information, yet all contribute to a unified perceptual inference.

\subsection{The Look-up Table as Compressed Representations}

The look-up table $\mathbf{D}_m$ provides a computational abstraction of how the brain creates compressed representations. Higher cortical areas receive compressed information through convergent projections; the look-up table captures essential information needed for inference at higher levels, establishing a Markov property where higher layers depend on lower layers only through summary statistics.

\subsection{Parallel Processing Streams and Integration}

While parallel MCMC distributes computations within a single model, RGP can be extended to model multiple distinct processing streams, such as the dorsal and ventral streams in visual cortex \citep{livingstone1988, goodale1992}. Different streams can be modeled as separate hierarchical pathways sharing the same GP structure but with potentially different correlation parameters. Integration across streams occurs at higher levels through precision-weighted combination, analogous to Bayesian cue combination \citep{ernst2002, knill2004}.

\subsection{Summary}

The parallel computation architecture of RGP provides a compelling analog of neurobiological processing at multiple levels. At the computational level, parallel MCMC distributes calculations within a single hierarchical model, with local update rules mirroring synaptic plasticity. At the architectural level, RGP can be extended to model multiple parallel processing streams, with integration through precision-weighted combination.

\section{Discussion}
\label{sec:discussion}

\subsection{Implications for Understanding Cortical Computation}

The mapping between RGP components and neuronal populations provides a concrete hypothesis about how hierarchical Bayesian inference could be implemented in cortex. This hypothesis is consistent with a wide range of anatomical, physiological, and behavioral evidence, and it generates testable predictions that can guide future experiments.

One key implication is that cortical computation is fundamentally probabilistic and involves shared constraints across areas. The brain does not compute deterministic functions of sensory input but maintains probability distributions over hidden causes and updates these distributions using Bayesian principles. Moreover, different cortical areas are not independent but share common computational principles arising from shared developmental and genetic constraints, while being able to differentiate through experience. The RGP's single GP with learnable layer correlation captures this balance between sharing and specialization.

A second implication is that hierarchical structure is essential for flexible, context-sensitive processing. The learnable layer dependence in RGP allows each cortical area to maintain its own specialization while still benefiting from shared structure, enabling the emergence of qualitatively different representations at each hierarchical stage. This contrasts with both simple feedforward architectures and models with independent layers. The brain's hierarchical organization, with its rich pattern of feedforward and feedback connections, reflects the need for both specialization and integration.

A third implication is that dynamics matter. The brain is not a static feedforward processor but a recurrent dynamical system that iteratively refines its representations through reciprocal interactions. The MCMC dynamics in RGP capture this iterative refinement, with each pass through the hierarchy corresponding to a step in the sampling process. The time course of neural responses---from initial transient to sustained activity---reflects the convergence of this sampling process to a stable posterior.

A fourth implication, highlighted by our analysis of parallel computation in Section \ref{sec:parallel-computation}, is that the brain's massive parallelism is not merely an implementation detail but a computational necessity. The parallel architecture enables simultaneous processing across multiple data points and layers, allowing for efficient inference in complex environments. The RGP framework demonstrates how such parallelism can be reconciled with principled Bayesian inference through local update rules and compressed representations.

\subsection{Implications for NeuroAI}

The NeuroAI research program seeks to build bridges between neuroscience and artificial intelligence, using insights from each field to advance the other \citep{zunger2024, hassabis2017}. The RGP framework contributes to this program in several important ways.

For neuroscience, RGP provides a formal language for expressing and testing hypotheses about cortical computation. The mapping between model components and neuronal populations generates precise predictions about neural activity that can be tested experimentally. The learnable layer dependence captured by $r_{1g}$ predicts that the functional connectivity between cortical areas should be dynamically modulated by task demands, a phenomenon that has been observed empirically \citep{buchel1999, friston1997}. The variational free energy formulation links these predictions to fundamental principles of self-organization, providing a normative account of why cortical circuits are organized as they are. The parallel computation analysis further reveals how the brain's distributed architecture can implement efficient Bayesian inference through local interactions.

For artificial intelligence, RGP offers architectural principles that could inspire more robust and interpretable deep learning systems. The single GP with layer-index input prevents representational collapse while allowing learnable sharing across layers, enabling deeper networks that maintain expressivity. The spike-and-slab variable selection provides a principled approach to attention and gating that could be incorporated into transformer architectures. The Bayesian foundations offer a path toward networks that provide reliable uncertainty estimates, addressing a key limitation of current deep learning systems \citep{gal2016}. Moreover, the parallel MCMC implementation demonstrates how scalable Bayesian inference can be achieved through distributed computation---a lesson directly applicable to large-scale AI systems.

The contrast with traditional deep learning models and Bayesian deep learning approaches highlighted in Sections \ref{sec:comparison-traditional} and \ref{sec:comparison-bayesian} underscores the value of biologically inspired architectures. While conventional methods have achieved remarkable engineering successes, their divergence from predictive coding principles limits their utility as models of brain function. The RGP framework, by incorporating neurobiological constraints such as shared structure with learnable specialization, offers a more faithful account of cortical computation while potentially also yielding practical benefits in terms of uncertainty quantification, interpretability, and parallel scalability.

\subsection{Future Directions}

Several important directions for future research emerge from this synthesis. First, the computational demands of RGP inference remain substantial. While parallel MCMC achieves scalability to moderate datasets and depths, the hours required for simulation would be prohibitive for many applications without sufficient computational resources. Developing more efficient inference algorithms---perhaps drawing on variational methods or amortized inference---could make RGP more widely applicable. The recent success of variational inference for deep GPs \citep{salimbeni2017} suggests promising directions. Additionally, exploring analog computing architectures that more closely mimic neural dynamics could yield orders-of-magnitude speedups \citep{hennequin2017}.

Second, the current RGP formulation assumes a fixed hierarchical structure with predetermined depth. A more flexible model would allow the hierarchical structure itself to be learned from data, with the number of layers and the connectivity between layers adapting to task demands. This could be achieved through nonparametric priors on the number of layers or through constructive algorithms that grow the network as needed. Such models would better capture the brain's remarkable plasticity and its ability to reorganize in response to experience \citep{buonomano2009}.

Third, the relationship between RGP inference and neural dynamics deserves deeper exploration. The MCMC sampling we employ operates on timescales vastly longer than neural dynamics, but it may be possible to reinterpret the iterative updates as corresponding to the settling dynamics of a recurrent neural network. Recent work on predictive coding networks has shown that inference learning can be implemented through local dynamics that converge to the same fixed points as variational inference \citep{millidge2020}. Exploring whether similar dynamics can be derived for RGP models would strengthen the biological plausibility of the framework. The parallel computation analysis in Section \ref{sec:parallel-computation} provides a foundation for such investigations.

Fourth, the current RGP framework focuses on supervised learning, while the brain operates in a largely unsupervised manner. Extending the model to unsupervised settings---perhaps by treating all layers as latent and learning to generate sensory data from the top down---would bring it closer to predictive coding formulations that emphasize generative modeling \citep{friston2005}. This extension could also incorporate the kind of structure learning discussed by \citet{rutar2022}, where agents must infer not only the parameters of a known model but the model structure itself.

Fifth, empirical testing of the framework's predictions is essential. The laminar-specific predictions outlined in Section \ref{subsec:laminar_dynamics} could be tested with existing techniques, providing evidence for or against the RGP-based account of predictive coding. The prediction about learnable layer dependence---that the functional connectivity between areas should be task-modulated---could be tested through fMRI or electrophysiological studies that manipulate task demands. The parallel computation predictions---including the relationship between local update rules and synaptic plasticity, and the role of compressed representations in hierarchical inference---could be tested through combinations of electrophysiology, imaging, and behavioral experiments. Collaboration between computational modelers and experimental neuroscientists will be crucial for refining the model and connecting it to empirical data.

\section{Conclusion}
\label{sec:conclusion}

The Recursive Gaussian Process framework represents a significant advance in Bayesian deep learning, offering a principled solution to degeneracy problems while maintaining rigorous uncertainty quantification. The key innovation---using a single Gaussian process $g(\cdot,\cdot)$ taking both layer index and input value as arguments, with learnable correlation across layers controlled by $r_{1g}$---provides a mathematically elegant way to balance sharing and specialization in hierarchical models.

We have argued that RGPs provide a compelling computational model of predictive coding, with connections that are not merely analogical but mathematical and neurobiological. The single GP corresponds to shared genetic and developmental constraints across cortical areas; $r_{1g}$ captures experience-dependent differentiation; recursive composition of layer-specific evaluations corresponds to hierarchical message passing; uncertainty propagation corresponds to precision-weighted prediction error coding; MCMC dynamics correspond to recurrent interactions; parallel computation mirrors distributed processing.

The mapping onto the canonical microcircuit provides a concrete neurobiological implementation of these computational principles. The free energy principle shows that RGP inference minimizes variational free energy, connecting algorithmic description to computational imperative. The epistemological analysis of \citet{colombo2012} reminds us that establishing a realist interpretation requires bridging behavioral and mechanistic evidence---a project to which RGP contributes through detailed neuronal mapping and testable predictions.

The brain remains the most sophisticated inference machine we know. The RGP framework, grounded in predictive coding and the canonical microcircuit, offers a promising approach to meeting the challenge of understanding its computational principles---one that respects both the complexity of the task and the biological reality of the substrate.

\section*{Acknowledgments}
%\textcolor{purple}
%MD and SB  thank DR for his valuable insights that introduced us to predictive coding. 
We acknowledge the use of DeepSeek for grammatical and stylistic improvements to the manuscript; the AI tool played no role in the intellectual content or conclusions 
of this work.

%\clearpage
\bibliographystyle{plainnat}
\bibliography{rgp_refs}

\end{document}